\documentclass[%
  twocolumn,
 showpacs,
 showkeys,
 preprintnumbers,
 amsmath,amssymb,
 aps,
  pra,
  longbibliography,
 ]{revtex4-1}

\usepackage[dvipsnames]{xcolor}

\usepackage{mathptmx}

\usepackage{amssymb,amsthm,amsmath}

\usepackage{tikz}
\usetikzlibrary{calc}

\usepackage[breaklinks=true,colorlinks=true,anchorcolor=blue,citecolor=blue,filecolor=blue,menucolor=blue,pagecolor=blue,urlcolor=blue,linkcolor=blue]{hyperref}
\usepackage{graphicx}
\usepackage{url}

\usepackage{xcolor}
\usepackage{colortbl}

\begin{document}

\title{Roots and (re)sources of value (in)definiteness {\it versus} contextuality.\\
A contribution to the Pitowsky Volume in memory of Itamar Pitowsky (1950--2010)}

\author{Karl Svozil}
\email{svozil@tuwien.ac.at}
\homepage{http://tph.tuwien.ac.at/~svozil}

\affiliation{Institute for Theoretical Physics,
Vienna  University of Technology,
Wiedner Hauptstrasse 8-10/136,
1040 Vienna,  Austria}

\date{\today}

\begin{abstract}
In Itamar Pitowsky's reading of the Gleason and the Kochen-Specker theorems, in particular, his Logical Indeterminacy Principle, the emphasis is on the {\em value indefiniteness} of observables which are not within the preparation context. This is in stark contrast to the prevalent term {\em contextuality} used by many researchers in informal, heuristic yet omni-realistic and potentially misleading ways. This paper discusses both concepts and argues in favor of value indefiniteness in all but a continuum of contexts intertwining in the vector representing a single pure (prepared) state. Even more restrictively, and inspired by operationalism but not justified by Pitowsky's Logical Indeterminacy Principle or similar, one could identify with a ``quantum state'' a single quantum context -- aka the respective maximal observable, or, in terms of its spectral decomposition, the associated orthonormal basis -- from the continuum of intertwining context, as per the associated maximal observable actually or implicitly measured.
\end{abstract}

\keywords{Value indefiniteness, Pitowsky's Logical Indeterminacy Principle, Quantum mechanics, Gleason theorem, Kochen-Specker theorem, Born rule}

\maketitle

\section{Introduction}

An upfront {\it caveat} seems in order: The following is a rather subjective narrative of my reading
of Itamar Pitowsky's thoughts about classical value indeterminacy on quantum logical structures of observables,
amalgamated with my current thinking on related issues.
I have never discussed these matters with Itamar Pitovsky
explicitly; therefore the term ``my reading'' should be taken rather literally; namely as taken
from his publications.
In what follows classical value indefiniteness on collections of (intertwined) quantum observables
will be considered a consequence, or even a synonym, of what he called indeterminacy.
Whether or not this identification is justified is certainly negotiable; but in what follows
this is taken for granted.

The term value indefiniteness has been stimulated by recursion theory~\cite{rogers1,odi:89,Smullyan1993-SMURTF},
and in particular by {\em partial functions}~\cite{Kleene1936} --
indeed the notion of partiality has not diffused into physical theory formation,
and might even appear alien to the very notion of functional value assignments --
and yet it appears to be necessary~\cite{2012-incomput-proofsCJ,PhysRevA.89.032109,2015-AnalyticKS}
if one insists (somewhat superficially)
on classical interpretations of quantized systems.

Value indefiniteness/indeterminacy will be contrasted with some related interpretations and approaches,
in particular, with contextuality.
Indeed, I believe that contextuality was rather foreign to Itamar Pitowsky's thinking:
the term {\em ``contextuality''} appears marginally --
as in ``a different context'' -- in his book
{\em Quantum Probability - Quantum Logic}~\cite{pitowsky},
nowhere in his reviews on Boole-Bell type inequalities~\cite{pitowsky-89a,Pit-94},
and mostly with reference to {\em contextual quantum probabilities} in his late writings~\cite{pitowsky-06}.
The emphasis on value indefiniteness/indeterminacy was, I believe, independently shared
by Asher Peres as well as Ernst Specker.

I met Itamar Pitowsky~\cite{BUB201085} personally rather late;
after he gave a lecture entitled {\em ``All Bell Inequalities''} in Vienna~\cite{ESI-AR-2000} on September 6th, 2000.
Subsequent discussions resulted in a joint paper~\cite{2000-poly}
(stimulating further research~\cite{sliwa-2003,collins-gisin-2003}).
It presents an application of his correlation polytope method~\cite{pitowsky-86,pitowsky,pitowsky-89a,Pit-91,Pit-94}
to more general configurations than had been studied before.
Thereby semi-automated symbolic as well as numeric computations have been used.

Nevertheless, the violations of what Boole called~\cite[p.~229]{Boole-62} {\em ``conditions of possible experience,''}
obtained through solving the hull problem of classical correlation polytopes, was just one route
to quantum indeterminacy pursued by Itamar Pitowsky.
One could identify at least two more passages he contributed to:
One approach~\cite{Pitowsky2003395,pitowsky-06} compares differences of classical with quantum predictions
through conditions and constraints imposed by certain intertwined configurations of observables which I like to call {\em quantum clouds}~\cite{svozil-2018-c}.
And another approach~\cite{pitowsky:218,hru-pit-2003} pushes these predictions to the limit of logical inconsistency;
such that any attempt of a classical description fails relative to the assumptions.
In what follows we shall follow all three pursuits and relate them to new findings.

\section{Stochastic value indefiniteness/indeterminacy by Boole-Bell type conditions of possible experience}

The basic idea to obtain all classical predictions -- including classical probabilities, expectations as well as consistency constraints thereof  --
associated with (mostly complementary; that is, non-simultaneously measurable) collections of observables
is quite straightforward:
Figure out all ``extreme'' cases or states which would be classically allowed.
Then construct all classically conceivable situations by forming suitable combinations of the former.

Formally this amounts to performing the following steps~\cite{pitowsky-86,pitowsky,pitowsky-89a,Pit-91,Pit-94}:
\begin{itemize}
\item
Contemplate about some concrete structure of observables and their interconnections in intertwining observables -- the quantum cloud.

\item
Find all two-valued states of that quantum cloud.
(In the case of ``contextual inequalities''~\cite{cabello:210401}
include all variations of true/1 and false/0, irrespective of exclusivity;
thereby often violating the Kolmogorovian axioms of probability theory even within a single context.)

\item
Depending on one's preferences, form all (joint) probabilities and expectations.

\item
For each of these two-valued states, evaluate the
joint  probabilities and expectations  as  products of the single particle probabilities and expectations they are formed of
(this reflects statistical independence of the constituent observables).

\item
For each of the two-valued states,
form a tuple containing these relevant (joint) probabilities and expectations.

\item
Interpret this tuple as a vector.

\item
Consider the set of all such vectors -- there are as many as there are two-valued states, and their dimension depends on the number of
(joint) probabilities and expectations considered -- and interpret them as vertices forming a convex polytope.

\item
The convex combination of all conceivable two-valued states yields the surface of this polytope;
such that every point inside its convex hull corresponds to a classical probability distribution.

\item
Determine the conditions of possible experience
by solving the hull problem -- that is, by computing the hyperplanes which determine the inside--versus--outside criteria for that polytope.
These then can serve as necessary criteria for all classical probabilities and expectations considered.

\end{itemize}

The systematic application of this method yields necessary criteria for classical probabilities and expectations
which are violated by the quantum probabilities and expectations.
Since I have reviewed this subject exhaustively~\cite[Sect.~12.9]{svozil-2016-pu-book} (see also Ref.~\cite{svozil-2017-b})
I have just sketched it to obtain a taste for its relevance for quantum indeterminacy.
As is often the case in mathematical physics
the method seems to have been envisioned independently a couple of times.
From its (to the best of my knowledge) inception by Boole~\cite{Boole-62}
it has been discussed in the measure theoretic context by Chochet theory~\cite{Bishop-Leeuw-1959}
and by Vorobev~\cite{Vorobev-1962}.
Froissart~\cite{froissart-81,cirelson} might have been the first explicitly proposing it as a method to generalized Bell-type inequalities.
I suggested its usefulness for non-Boolean cases~\cite{svozil-2001-cesena} with ``enough'' two-valued states; preferable sufficiently many to
allow a proper distinction/separation of all observables (cf. Kochen and Specker's Theorem~0~\cite[p.~67]{kochen1}).
Consideration of the pentagon/pentagram logic -- that is, five cyclically intertwined contexts/blocks/Boolean subalgebras/cliques/orthonormal bases
popularized the subject and also rendered new predictions which could be used to differentiate
classical from quantized systems~\cite{Klyachko-2002,Klyachko-2008,Bub-2009,Bub-2010,Badziag-2011}.

A {\it caveat:}
the  obtained criteria involve multiple mutually complementary summands which are not all simultaneously measurable.
Therefore, different terms,
when evaluated experimentaly,
correspond to different, complementary measurement configurations.
They are obtained at different times and on different particles and samples.

Explicit, worked examples can, for instance, be found in Pitowsky's book~\cite[Section~2.1]{pitowsky},
or papers~\cite{Pit-94} (see also Froissart's example~\cite{froissart-81}).
Empirical findings are too numerous to even attempt a just appreciation of all the efforts that went into
testing classicality. There is overwhelming evidence that the quantum predictions are correct;
and that they violate Boole's
conditions of possible classical experience~\cite{clauser-talkvie} relative to the assumptions (basically non-contextual realism and locality).

So, if Boole's conditions of possible experience are violated, then
they can no longer be considered appropriate for any reasonable ontology forcing "reality" upon them.
This includes the realistic~\cite{stace} existence of hypothetical counterfactual observables: ``unperformed experiments seem to have no
consistent outcomes''~\cite{peres222}.
The inconsistency of counterfactuals (in Specker's scholastic terminology {\it infuturabilities}~\cite{specker-60,specker-ep})
provides a connection to value indefiniteness/indeterminacy -- at least, and let me again repeat earlier provisos, relative to the assumptions.
More of this, piled higher and deeper, has been supplied
by Itamar Pitowsky, as will be discussed later.

\section{Interlude: quantum probabilities from Pythagorean ``views on vectors''}

Quantum probabilities are vector based.
At the same time those probabilities mimic ``classical'' ones whenever they must be classical;
that is, among mutually commuting observables which can be measured simultaneously/concurrently on the same particle(s) or samples
--
in particular, whenever those observables correspond to projection operators which are either orthogonal (exclusive)
or identical (inclusive).

At the same time, quantum probabilities appear ``contextual'' (I assume he had succumbed to the prevalent nomenclature at that late time)
according to Itamar Pitowsky's late late writings~\cite{pitowsky-06}
if they need not be classical: namely among non-commuting observables.
(The term ``needs not'' derives its justification from the finding that there exist
situations~\cite{e-f-moore,wright} involving complementary observables
with a classical probability interpretation~\cite{svozil-2001-eua}).

Thereby,
classical probability theory is maintained for simultaneously co-measurable
(that is, non-complementary) observables.
This essentially amounts to the validity of the Kolmogorov axioms of probability theory  of such observables within a given
context/block/Boolean subalgebra/clique/orthonormal basis, whereby the probability of an event associated with an observable
\begin{itemize}
\item
is a non-negative real number between $0$ and $1$;
\item
is $1$ for an event associated with an observable occurring with certainty (in particular, by considering any observable or its complement);
as well as
\item
additivity of probabilities for events associated with mutually exclusive observables.
\end{itemize}

Sufficiency is assured by an elementary geometric argument~\cite{Gleason} which is based upon the Pythagorean theorem;
and which can be used to explicitly construct vector-based probabilities satisfying the aforementioned Kolmogorov axioms within contexts:
Suppose a pure state of a quantized system is formalized
by the unit state vector $\vert \psi \rangle$.
Consider some orthonormal basis ${\cal B} = \{ \vert {\bf e}_1\rangle ,\ldots , \vert {\bf e}_n\rangle \}$ of $\cal V$.
Then the  square
$P_\psi({\bf e}_i)= \vert \langle \psi \vert {\bf e}_i\rangle \vert^2$
of the length/norm
$\sqrt{
\langle \psi \vert {\bf e}_i\rangle
\langle {\bf e}_i \vert \psi\rangle
}$
of the orthogonal projection
$\left(\langle \psi \vert {\bf e}_i\rangle  \right)  \vert {\bf e}_i\rangle$
of that unit vector
$\vert \psi \rangle$
along the basis element
$\vert {\bf e}_i\rangle$
can be interpreted as
the probability of the event associated with the $0-1$-observable (proposition) associated with the basis vector  $\vert {\bf e}_i\rangle$
(or rather the orthogonal projector $\textsf{\textbf{E}}_i =  \vert {\bf e}_i \rangle \langle   {\bf e}_i  \vert $
associated with the dyadic product of the basis vector $\vert {\bf e}_i\rangle$);
given a quantized physical system which has been prepared to be in a pure state
$\vert \psi \rangle $.
Evidently, $1\le P_\psi({\bf e}_i)\le1$,
and $\sum_{i=1}^n P_\psi({\bf e}_i)=1$.
In that Pythagorean way, every context, formalized by an orthonormal basis ${\cal B}$,
``grants a (probabilistic) view'' on the pure state $\vert \psi \rangle$.

It can be expected that these Pythagorean-style probabilities are different from classical probabilities
almost everywhere --
that is, for almost all relative measurement positions.
Indeed, for instance, whereas classical two-partite correlations are
linear in the relative measurement angles, their respective quantum correlations follow trigonometric functions
-- in particular, the cosine for ``singlets''~\cite{peres}.
These differences, or rather the vector-based Pythagorean-style quantum  probabilities, are the ``root cause'' for violations of
Boole's aforementioned conditions of possible experience in quantum setups.

Because of the convex combinations from which they are derived, all of these conditions of possible experience contain only
{\em linear}
constraints~\cite{Boole,Boole-62,Frechet1935,Hailperin-1965,Hailperin-86,Ursic1984,Ursic:1986:GFL:3023712.3023752,Ursic1988,Beltrametti-1991,Pykacz-1991,Pulmannova-1992,Beltrametti-1993,Beltrametti-1994,DvurLaen-1994,Beltrametti1995,Beltrametti-1995,Noce-1995,Laenger1995,DvurLaen-1995,DvurLaen-1995b,Beltrametti-1996,Pulmannova-2002}.
And because linear combinations of linear operators remain linear,
one can identify the terms occurring in conditions of possible experience with linear self-adjoint operators,
whose sum yields a self-adjoint operator, which stands for the ``quantum version'' of the respective conditions of possible experience.
This operator has a spectral decomposition whose min-max eigenvalues correspond to the quantum bounds~\cite{filipp-svo-04-qpoly,filipp-svo-04-qpoly-prl},
which thereby generalize the  Tsirelson bound~\cite{cirelson:80}.
In that way, every condition of possible experience which is violated by the quantum probabilities provides a direct criterium for non-classicality.


\section{Classical value indefiniteness/indeterminacy by direct observation}

In addition to the ``fragmented, explosion view''
criteria allowing ``nonlocality'' {\it via} Einstein separability~\cite{wjswz-98} among its parts,
classical predictions from quantum clouds -- essentially intertwined (therefore the Hilbert space dimensionality has to be greater than two)
arrangements of contexts -- can be used as a criterium for quantum ``supremacy'' over (or rather ``otherness''
or ``distinctiveness'' from) classical predictions.
Thereby it is sufficient to observe of a single outcome of a quantized system which directly contradicts the classical predictions.

\begin{figure}
\begin{center}
\begin{tikzpicture}  [scale=0.25, rotate=0]

\newdimen\ms
\ms=0.05cm

\tikzstyle{every path}=[line width=1pt]

\tikzstyle{c3}=[circle,inner sep={\ms/8},minimum size=3*\ms]
\tikzstyle{c2}=[circle,inner sep={\ms/8},minimum size=1.7*\ms]
\tikzstyle{c1}=[circle,inner sep={\ms/8},minimum size=0.8*\ms]

\newdimen\R
\R=6cm     



\path
  ({180 + 0 * 360 /6}:\R      ) coordinate(1)
  ({180 + 30 + 0 * 360 /6}:{\R * sqrt(3)/2}      ) coordinate(2)
  ({180 + 1 * 360 /6}:\R   ) coordinate(3)
  ({180 + 30 + 1 * 360 /6}:{\R * sqrt(3)/2}   ) coordinate(4)
  ({180 + 2 * 360 /6}:\R  ) coordinate(5)
  ({180 + 30 + 2 * 360 /6}:{\R * sqrt(3)/2}  ) coordinate(6)
  ({180 + 3 * 360 /6}:\R  ) coordinate(7)
  ({180 + 30 + 3 * 360 /6}:{\R * sqrt(3)/2}  ) coordinate(8)
  ({180 + 4 * 360 /6}:\R     ) coordinate(9)
  ({180 + 30 + 4 * 360 /6}:{\R * sqrt(3)/2}     ) coordinate(10)
  ({180 + 5 * 360 /6}:\R     ) coordinate(11)
  ({180 + 30 + 5 * 360 /6}:{\R * sqrt(3)/2}     ) coordinate(12)
  (0:0     ) coordinate(13)
;


\draw [color=orange] (1) -- (2) -- (3);
\draw [color=red] (3) -- (4) -- (5);
\draw [color=green] (5) -- (6) -- (7);
\draw [color=blue] (7) -- (8) -- (9);
\draw [color=magenta] (9) -- (10) -- (11);    %
\draw [color=olive] (11) -- (12) -- (1);    %
\draw [color=lime] (4) -- (13) -- (10);    %

%
%
\draw (1) coordinate[minimum size=1cm];  %
\draw (1) coordinate[c3,fill=red,label={left:\scriptsize $\{ 1,2,3 \} $}];   %
\draw (1) coordinate[c2,fill=olive];  %
\draw (2) coordinate[c3,fill=orange,label={left:\scriptsize $\{ 4,5,6,7,8,9 \}$}];    %
\draw (3) coordinate[c3,fill=red,label={left:\scriptsize $\{ 10,11,12,13,14 \} $}]; %
\draw (3) coordinate[c2,fill=orange];  %
\draw (4) coordinate[c3,fill=red,label={below:\scriptsize $\{ 2,6,7,8 \}$}];  %
\draw (4) coordinate[c2,fill=lime];  %
\draw (5) coordinate[c3,fill=green,label={right:\scriptsize $\{ 1,3,4,5,9 \} $}];  %
\draw (5) coordinate[c2,fill=red];  %
\draw (6) coordinate[c3,fill=green,label={right:\scriptsize $\{ 2,6,8,11,12,14 \} $}];
%
\draw (7) coordinate[c3,fill=blue,label={right:\scriptsize $\{ 7,10,13 \}$}];  %
\draw (7) coordinate[c2,fill=green];  %
\draw (8) coordinate[c3,fill=blue,label={right:\scriptsize $\{ 3,5,8,9,11,14 \}$}];  %
\draw (9) coordinate[c3,fill=magenta,label={right:\scriptsize $\{ 1,2,4,6,12 \}$}];
\draw (9) coordinate[c2,fill=blue];  %
\draw (10) coordinate[c3,fill=magenta,label={above:\scriptsize $\{ 3,9,13,14 \}$}];  %
\draw (10) coordinate[c2,fill=lime];  %
\draw (11) coordinate[c3,fill=olive,label={left:\scriptsize $\{ 5,7,8,10,11 \}$}];
\draw (11) coordinate[c2,fill=blue];  %
\draw (12) coordinate[c3,fill=olive,label={left:\scriptsize $\{ 4,6,9,12,13,14 \}$}];  %
%
\draw (13) coordinate[c3,fill=lime,label={[xshift=-0.5mm]above right:\scriptsize $\{1,4,$}];  %
\draw (13) coordinate[c3,fill=lime,label={[xshift=-0.5mm]right:\scriptsize $5,10,$}];  %
\draw (13) coordinate[c3,fill=lime,label={[xshift=-0.5mm]below right:\scriptsize $11,12\}$}];  %
\end{tikzpicture}
\end{center}
\caption{The convex structure of classical probabilities in this (Greechie) orthogonality diagram
representation of the Specker bug quantum or partition logic
is reflected in its partition logic, obtained through indexing all 14 two-valued measures, and adding an index $1\le i \le 14$
if the $i$th two-valued measure is 1 on the respective atom.
Concentrate on the outermost left and right observables, depicted by squares:
Positivity and convexity requires that $0\le \lambda_i\le1$ and
$\lambda_1+\lambda_2+\lambda_3+ \lambda_7+\lambda_{10}+\lambda_{13} \le
\sum_{i=1}^{14} \lambda_i =1$.
Therefore, if a classical system is prepared (a generalized urn model/automaton logic is ``loaded'')
such that $\lambda_1+\lambda_2+\lambda_3=1$,
then  $\lambda_7+\lambda_{10}+\lambda_{13}=0$, which results in a TIFS:
the classical prediction is that the latter outcome never occurs
if the former preparation is certain.
}
\label{2018-pit-f1}
\end{figure}
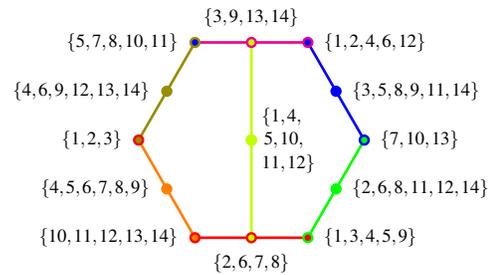

One example of such a configuration of quantum observables forcing a ``one-zero rule"~\cite{svozil-2006-omni} because of a
true-implies-false set of two-valued classical states (TIFS)~\cite{2018-minimalYIYS}
is the ``Specker bug'' logic~\cite[Fig.~1, p.~182]{kochen2}
called  ``cat's cradle''~\cite{Pitowsky2003395,pitowsky-06} by Itamar Pitowsky
(see also Refs.~\cite[Fig.~B.l. p.~64]{Belinfante-73}, \cite[p.~588-589]{stairs83},
\cite[Sects.~IV, Fig.~2]{clifton-93}  and \cite[p.~39, Fig.~2.4.6]{pulmannova-91} for early discussions), as
depicted in Fig.~\ref{2018-pit-f1}.

For such configurations, it is often convenient to represent both its labels as well as
the classical probability distributions in terms of a partition logic~\cite{svozil-2001-eua}
of the set of two-valued states -- in this case, there are 14 such classical states. Every maximal observable is characterized by a context.
The atoms of this context are labeled according to the indices of the two-valued measure with the value $1$ on this atom.
The axioms of probability theory require that, for each two-valued state,
and within each context,
there is exactly one such atom.
As a result, as long as the set of two-valued states is separating~\cite[Theorem~0]{kochen1},
one obtains a set of partitions of the set of two-valued states; each partition corresponding to a context.

Classically, if one prepares the system to be in the state $\{1,2,3\}$ -- standing for any one of the classical two-valued states $1$, $2$ or $3$
or their convex combinations -- then there is no chance that the ``remote'' target state $\{ 7,10,13 \}$ can be observed.
A direct observation of  quantum ``supremacy''
(or rather superiority in terms of the frequencies predicted with respect to classical frequencies) is then possibly by
some faithful orthogonal representation (FOR)~\cite{lovasz-89,Parsons-1989,Cabello-2010-ncoptaa,Portillo-2015}
of this graph. In the particular Specker but/cats cradle configuration,
an elementary geometric argument~\cite{cabello-1994,Cabello-1996-diss}
forces the relative angle between the quantum states
$\vert \{1,2,3\} \rangle$
and
$\vert \{7,10,13\} \rangle$
in three dimensions to be not smaller than $\text{arctan} \left(2\sqrt{2}\right)$,
so that the quantum prediction of the occurrence of the event associated with state $\vert \{7,10,13\} \rangle$,
if the system was prepared in state
$\vert \{1,2,3\} \rangle$
is that the probability can be at most
$\vert \langle \{1,2,3\} \vert \{7,10,13\}  \rangle \vert^2 =
\cos^2 \left[
\text{arctan} \left(2\sqrt{2}\right)
\right]
=\frac{1}{9}$.
That is, on the average, if the system was prepared in
state
$\vert \{1,2,3\} \rangle$
at most one of $9$ outcomes indicates that the system has the property associated with the
observable
 $\vert \{7,10,13\} \rangle \langle \vert \{7,10,13\}\vert$.
The occurrence of a single such event indicates quantum ``supremacy'' over the classical prediction of non-occurrence.

This limitation is only true for the particular quantum cloud involved.
Similar arguments with different quantum clouds resulting in TIFS
can be extended to arbitrary small relative angles between preparation and measurement states,
so that the relative quantum ``supremacy''
can be made arbitrarily high~\cite{2015-AnalyticKS}.
Classical value indefiniteness/indeterminacy comes naturally:
because -- at least relative to the assumptions regarding non-contextual value definiteness
of truth assignments, in particular, of intertwining, observables --
the existence of such definite values would enforce
non-occurrence of outcomes which are nevertheless observed in quantized systems.

Very similar arguments against classical value definiteness can be inferred from quantum clouds with true-implies-true
sets of two-valued
states (TITS)~\cite{stairs83,clifton-93,Johansen-1994,Vermaas-1994,Belinfante-73,Pitowsky-1982-subs,Hardy-92,Hardy-93,hardy-97,Cabello-1995-ppks,cabello-96,cabello-97-nhvp,Badziag-2011,Cabello-2013-HP,Cabello-2013-Hardylike,2018-minimalYIYS}.
There the quantum ``supremacy'' is in the non-occurrence of outcomes which classical predictions mandate to occur.

\section{Classical value indefiniteness/indeterminacy piled higher and deeper: The Logical Indeterminacy Principle}

For the next and final stage of classical value indefiniteness/indeterminacy on quantum clouds (relative to the assumptions)
one can combine two logics with simultaneous classical TIFS and TITS properties at the same terminals.
That is,
suppose one is preparing the same ``initial'' state, and measuring the same ``target'' observable;
nevertheless, contemplating
the simultaneous counterfactual existence of
two different quantum clouds of intertwined contexts
interconnecting those fixated ``initial'' state and measured ``target'' observable.
Whenever one cloud has the
TIFS and another cloud the TITS property (at the same terminals),
those quantum clouds induce contradicting classical predictions.
In such a setup the only consistent choice (relative to the assumptions; in particular, omni-existence and context independence)
is to abandon classical value definiteness/determinacy.
Because the assumption of classical value definiteness/determinacy for any such logic,
therefore, yields a complete contradiction,
thereby eliminating prospects for hidden
variable models~\cite{2012-incomput-proofsCJ,2015-AnalyticKS,svozil-2018-c}
satisfying the assumptions.

Indeed,
suppose that a quantized system is prepared in some pure quantum state.
Then
Itamar Pitowsky's~\cite{pitowsky:218,hru-pit-2003} {\em indeterminacy principle}
states that -- relative to the assumptions;
in particular, global classical value definiteness for all observables involved, as well as context-independence of observables in which contexts intertwine --
any other distinct (non-collinear) observable which is not orthogonal can neither occur nor not occur.
This can be seen as an extension of both Gleason's theorem~\cite{Gleason,ZirlSchl-65} as well as the Kochen-Specker theorem~\cite{kochen1}
implying and utilizing the non-existence of any two-valued global truth assignments on even finite quantum clouds.

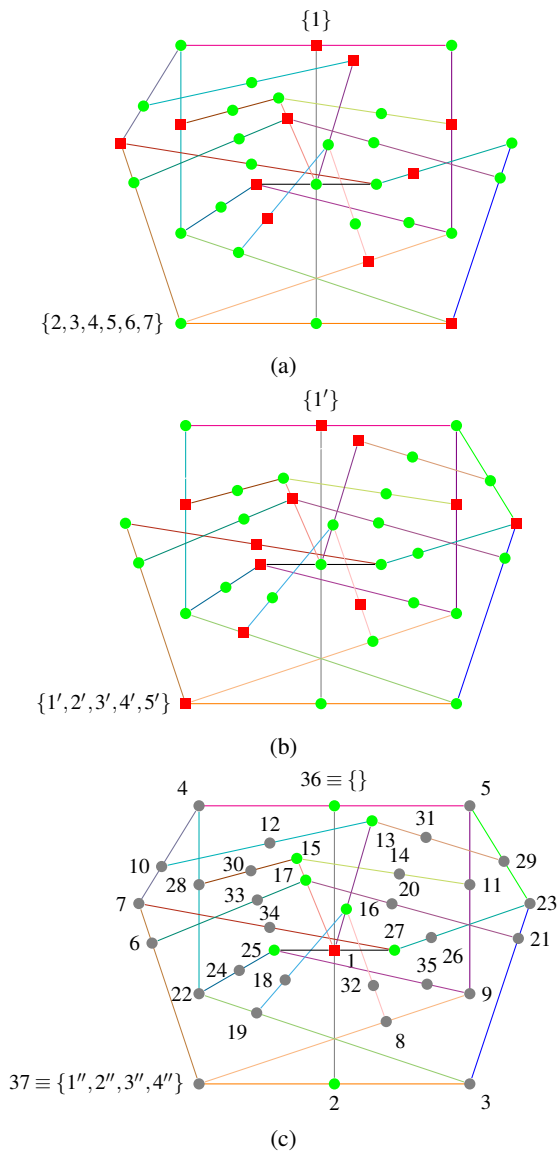
\begin{figure}
\begin{tabular}{c}
\newif\iflabel \labelfalse
\begin{tikzpicture}  [scale=0.20, rotate=0]

        \tikzstyle{c1}=[color=green,circle,inner sep=1.5]
        \tikzstyle{s1}=[color=red,rectangle,inner sep=2]
        \tikzstyle{l1}=[draw=none,circle,minimum size=3]


\draw [color=orange]   (4,0) coordinate[c1,fill,label=180:{\color{black}\footnotesize $\{2,3,4,5,6,7 \}$}] (b) -- (13,0)     coordinate[c1,fill,label=270:{\iflabel \tiny $P_2$\fi}] (2) -- (22,0)  coordinate[s1,fill,label=315:{\iflabel \tiny $P_3$\fi}] (3);
\draw [color=blue,   ] (3) -- (26,12)  coordinate[c1,fill,pos=0.8,label=0:{\iflabel \tiny $P_{21}$\fi}] (21) coordinate[c1,fill,label=0:{\iflabel \tiny $P_{23}$\fi}] (23);
\draw [color=white] (23) -- (22,18.5) coordinate[c1,fill,pos=0.4,color=white,label=0:{\iflabel \tiny $P_{29}$\fi}] (29) coordinate[c1,fill,label=45:{\iflabel \tiny $P_5$\fi}] (5);
\draw [color=magenta,] (5)-- (13,18.5)coordinate[s1,fill,label=90:{\color{black}\footnotesize $\{ 1 \}$}] (a) -- (4,18.5)  coordinate[c1,fill,label=135:{\iflabel \tiny $P_4$\fi}] (4);
\draw [color=CadetBlue, ] (4) -- (0,12)   coordinate[c1,fill,pos=0.6,label=180:{\iflabel \tiny $P_{10}$\fi}] (10) coordinate[s1,fill,label=180:{\iflabel \tiny $P_7$\fi}] (7);
\draw [color=brown,  ](7) -- (b)       coordinate[c1,fill,pos=0.2,label=90:{\iflabel \tiny $P_6$\fi}] (6);

        \draw [color=gray] (a) -- (2) coordinate[c1,fill,pos=0.5,label=315:{\iflabel \tiny $P_1$\fi}] (1);

        \draw [color=violet] (5) -- (22,6) coordinate[s1,fill,pos=0.4,label=0:{\iflabel \tiny $P_{11}$\fi}] (11) coordinate[c1,fill,label=0:{\iflabel \tiny $P_9$\fi}] (9);

\draw [color=Apricot] (9) -- (b) coordinate[s1,fill,pos=0.3,label=280:{\iflabel \tiny $P_8$\fi}] (8);

\draw [color=TealBlue] (4) -- (4,6) coordinate[s1,fill,pos=0.4,label=180:{\iflabel \tiny $P_{28}$\fi}] (28) coordinate[c1,fill,label=180:{\iflabel \tiny $P_{22}$\fi}] (22);
\draw [color=YellowGreen] (22) -- (3) coordinate[c1,fill,pos=0.2,label=260:{\iflabel \tiny $P_{19}$\fi}] (19);

        \coordinate (25) at ([xshift=-4cm]1);
        \coordinate (27) at ([xshift=4cm]1);

\draw [color=MidnightBlue]  (22) -- (25) coordinate[c1,fill,pos=0.5,label=115:{\iflabel \tiny $P_{24}$\fi}] (24) coordinate[s1,fill,label=270:{\iflabel \tiny $P_{25}$\fi}] (25);
\draw [color=Mulberry] (25) -- (9) coordinate[c1,fill,pos=0.8,label=90:{\iflabel \tiny $P_{35}$\fi}] (35);

\draw [color=BrickRed]  (7) -- (27) coordinate[c1,fill,pos=0.5,label=90:{\iflabel \tiny $P_{34}$\fi}] (34) coordinate[c1,fill,label=90:{\iflabel \tiny $P_{27}$\fi}] (27);
\draw [color=Emerald] (27) -- (23) coordinate[s1,fill,pos=0.25,label=270:{\iflabel \tiny $P_{26}$\fi}] (26);

\draw [color=BlueGreen]  (10) -- (15.5,17.5) coordinate[c1,fill,pos=0.5,label=90:{\iflabel \tiny $P_{12}$\fi}] (12) coordinate[s1,fill,label=15:{\iflabel \tiny $P_{13}$\fi}] (13);

\draw [color=RawSienna]  (28) -- (10.5,15) coordinate[c1,fill,pos=0.5,label=90:{\iflabel \tiny $P_{30}$\fi}] (30) coordinate[c1,fill,label=90:{\iflabel \tiny $P_{15}$\fi}] (15);
\draw [color=SpringGreen] (15) -- (11) coordinate[c1,fill,pos=0.6,label=90:{\iflabel \tiny $P_{14}$\fi}] (14);

\draw [color=Salmon]  (15) -- (1) coordinate[s1,fill,pos=0.2,label=15:{\iflabel \tiny $P_{17}$\fi}] (17);
\draw [color=Fuchsia] (1)-- (13) coordinate[c1,fill,pos=0.3,label=0:{\iflabel \tiny $P_{16}$\fi}] (16);

\draw [color=CornflowerBlue]  (19) -- (16) coordinate[s1,fill,pos=0.3,label=180:{\iflabel \tiny $P_{18}$\fi}] (18);
\draw [color=pink] (16) -- (8) coordinate[c1,fill,pos=0.7,label=180:{\iflabel \tiny $P_{32}$\fi}] (32);

\draw [color=PineGreen]  (6) -- (17) coordinate[c1,fill,pos=0.7,label=90:{\iflabel \tiny $P_{33}$\fi}] (33);
\draw [color=DarkOrchid] (17) -- (21) coordinate[c1,fill,pos=0.4,label=90:{\iflabel \tiny $P_{20}$\fi}] (20);

\draw [color=black] (25)  -- (1) -- (27);



\end{tikzpicture}
\\
(a)
\\
\newif\iflabel
\labelfalse
\begin{tikzpicture}  [scale=0.20, rotate=0]

        \tikzstyle{c1}=[color=green,circle,inner sep=1.5]
        \tikzstyle{s1}=[color=red,rectangle,inner sep=2]
        \tikzstyle{l1}=[draw=none,circle,minimum size=3]


\draw [color=orange] (4,0) coordinate[s1,fill,label=180:{\color{black}\footnotesize $\{ 1',2',3',4',5'\}$}] (b) -- (13,0)  coordinate[c1,fill,label=270:{\iflabel \tiny $P_2$\fi}] (2) -- (22,0) coordinate[c1,fill,label=315:{\iflabel \tiny $P_3$\fi}] (3);
\draw [color=blue, ] (3) -- (26,12) coordinate[c1,fill,pos=0.8,label=0:{\iflabel \tiny $P_{21}$\fi}] (21) coordinate[s1,fill,label=0:{\iflabel \tiny $P_{23}$\fi}] (23);
\draw [color=green] (23) -- (22,18.5) coordinate[c1,fill,pos=0.4,label=0:{\iflabel \tiny $P_{29}$\fi}] (29) coordinate[c1,fill,label=45:{\iflabel \tiny $P_5$\fi}] (5);
\draw [color=magenta,] (5)-- (13,18.5)coordinate[s1,fill,label=90:{\color{black}\footnotesize $\{ 1' \}$}] (a) -- (4,18.5) coordinate[c1,fill,label=135:{\iflabel \tiny $P_4$\fi}] (4);
\draw [color=white] (4) -- (0,12) coordinate[c1,color=white,fill,pos=0.6,label=180:{\iflabel \tiny $P_{10}$\fi}] (10) coordinate[c1,fill,label=180:{\iflabel \tiny $P_7$\fi}] (7);
\draw [color=brown, ] (7) -- (b)   coordinate[c1,fill,pos=0.2,label=90:{\iflabel \tiny $P_6$\fi}] (6);

   \draw [color=gray] (a) -- (2) coordinate[c1,fill,pos=0.5,label=315:{\iflabel \tiny $P_1$\fi}] (1);

   \draw [color=violet] (5) -- (22,6) coordinate[s1,fill,pos=0.4,label=0:{\iflabel \tiny $P_{11}$\fi}] (11) coordinate[c1,fill,label=0:{\iflabel \tiny $P_9$\fi}] (9);

\draw [color=Apricot] (9) -- (b) coordinate[c1,fill,pos=0.3,label=280:{\iflabel \tiny $P_8$\fi}] (8);

\draw [color=TealBlue] (4) -- (4,6) coordinate[s1,fill,pos=0.4,label=180:{\iflabel \tiny $P_{28}$\fi}] (28) coordinate[c1,fill,label=180:{\iflabel \tiny $P_{22}$\fi}] (22);
\draw [color=YellowGreen] (22) -- (3) coordinate[s1,fill,pos=0.2,label=260:{\iflabel \tiny $P_{19}$\fi}] (19);

   \coordinate (25) at ([xshift=-4cm]1);
   \coordinate (27) at ([xshift=4cm]1);

\draw [color=MidnightBlue] (22) -- (25) coordinate[c1,fill,pos=0.5,label=115:{\iflabel \tiny $P_{24}$\fi}] (24) coordinate[s1,fill,label=270:{\iflabel \tiny $P_{25}$\fi}] (25);
\draw [color=Mulberry] (25) -- (9) coordinate[c1,fill,pos=0.8,label=90:{\iflabel \tiny $P_{35}$\fi}] (35);

\draw [color=BrickRed] (7) -- (27) coordinate[s1,fill,pos=0.5,label=90:{\iflabel \tiny $P_{34}$\fi}] (34) coordinate[c1,fill,label=90:{\iflabel \tiny $P_{27}$\fi}] (27);
\draw [color=Emerald] (27) -- (23) coordinate[c1,fill,pos=0.25,label=270:{\iflabel \tiny $P_{26}$\fi}] (26);

\draw [color=white] (10) -- (15.5,17.5) coordinate[c1,color=white,fill,pos=0.5,label=90:{\iflabel \tiny $P_{12}$\fi}] (12) coordinate[s1,fill,label=15:{\iflabel \tiny $P_{13}$\fi}] (13);
\draw [color=Tan] (13) -- (29) coordinate[c1,fill,pos=0.4,label=90:{\iflabel \tiny $P_{31}$\fi}] (31);

\draw [color=RawSienna] (28) -- (10.5,15) coordinate[c1,fill,pos=0.5,label=90:{\iflabel \tiny $P_{30}$\fi}] (30) coordinate[c1,fill,label=90:{\iflabel \tiny $P_{15}$\fi}] (15);
\draw [color=SpringGreen] (15) -- (11) coordinate[c1,fill,pos=0.6,label=90:{\iflabel \tiny $P_{14}$\fi}] (14);

\draw [color=Salmon] (15) -- (1) coordinate[s1,fill,pos=0.2,label=15:{\iflabel \tiny $P_{17}$\fi}] (17);
\draw [color=Fuchsia] (1)-- (13) coordinate[c1,fill,pos=0.3,label=0:{\iflabel \tiny $P_{16}$\fi}] (16);

\draw [color=CornflowerBlue] (19) -- (16) coordinate[c1,fill,pos=0.3,label=180:{\iflabel \tiny $P_{18}$\fi}] (18);
\draw [color=pink] (16) -- (8) coordinate[s1,fill,pos=0.7,label=180:{\iflabel \tiny $P_{32}$\fi}] (32);

\draw [color=PineGreen] (6) -- (17) coordinate[c1,fill,pos=0.7,label=90:{\iflabel \tiny $P_{33}$\fi}] (33);
\draw [color=DarkOrchid] (17) -- (21) coordinate[c1,fill,pos=0.4,label=90:{\iflabel \tiny $P_{20}$\fi}] (20);

\draw [color=black] (25) -- (1) -- (27);

   \coordinate (ContextLabel) at ([shift=({-2cm,-3mm})]1);
   \draw (ContextLabel) coordinate[l1,label=90:{\iflabel \tiny $C_{26}$\fi}];

   \end{tikzpicture}
\\
(b)
\\
\begin{tikzpicture}  [scale=0.20, rotate=0]
\newif\iflabel \labelfalse
\labeltrue
       \tikzstyle{c1}=[color=gray,circle,inner sep=1.5]
       \tikzstyle{c2}=[color=green,circle,inner sep=1.5]
        \tikzstyle{s1}=[color=red,rectangle,inner sep=2]
        \tikzstyle{l1}=[draw=none,circle,minimum size=3]


\draw [color=orange]  (4,0)  coordinate[c1,fill,label=180:{\color{black}\footnotesize $37\equiv \{ 1'',2'',3'',4''\}$}] -- (13,0)    coordinate[c2,fill,label={[label distance=-1]270:{\iflabel \footnotesize \color{black}  $2$\fi}}] (2) -- (22,0)  coordinate[c1,fill,label={[label distance=-1]315:{\iflabel \footnotesize \color{black}  $3$\fi}}] (3);
\draw [color=blue] (3) -- (26,12)  coordinate[c1,fill,pos=0.8,label={[label distance=-1]0:{\iflabel \footnotesize \color{black}  ${21}$\fi}}] (21) coordinate[c1,fill,label={[label distance=-3]0:{\iflabel \footnotesize \color{black}  ${23}$\fi}}] (23);
\draw [color=green] (23) -- (22,18.5) coordinate[c1,fill,pos=0.4,label={[label distance=-1]0:{\iflabel \footnotesize \color{black}  ${29}$\fi}}] (29) coordinate[c1,fill,label={[label distance=-1]45:{\iflabel \footnotesize \color{black}  $5$\fi}}] (5);
\draw [color=magenta] (5)-- (13,18.5)coordinate[c2,fill,label=90:{\color{black}\footnotesize $36\equiv \{ \}$}] (a) -- (4,18.5)  coordinate[c1,fill,label={[label distance=-1]135:{\iflabel \footnotesize \color{black}  $4$\fi}}] (4);
\draw [color=CadetBlue] (4) -- (0,12)   coordinate[c1,fill,pos=0.6,label={[label distance=-1]180:{\iflabel \footnotesize \color{black}  ${10}$\fi}}] (10)  coordinate[c1,fill,label={[label distance=-1]180:{\iflabel \footnotesize \color{black}  $7$\fi}}] (7);
\draw [color=brown] (7) -- (b)      coordinate[c1,fill,pos=0.2,label={[label distance=-1]180:{\iflabel \footnotesize \color{black}  $6$\fi}}] (6);

        \draw [color=gray] (a) -- (2) coordinate[s1,fill,pos=0.52,label={[label distance=-1, yshift=2]357.5:{\iflabel \footnotesize \color{black}  $1$\fi}}] (1);

        \draw [color=violet] (5) -- (22,6) coordinate[c1,fill,pos=0.4,label={[label distance=-1]0:{\iflabel \footnotesize \color{black}  ${11}$\fi}}] (11) coordinate[c1,fill,label={[label distance=-1]0:{\iflabel \footnotesize \color{black}  $9$\fi}}] (9);

\draw [color=Apricot] (9) -- (b) coordinate[c1,fill,pos=0.3,label={[label distance=-1]280:{\iflabel \footnotesize \color{black}  $8$\fi}}] (8);

\draw [color=TealBlue] (4) -- (4,6) coordinate[c1,fill,pos=0.4,label={[label distance=-1]180:{\iflabel \footnotesize \color{black}  ${28}$\fi}}] (28) coordinate[c1,fill,label={[label distance=-3]180:{\iflabel \footnotesize \color{black}  ${22}$\fi}}] (22);
\draw [color=YellowGreen] (22) -- (3) coordinate[c1,fill,pos=0.2,label={[label distance=-1]260:{\iflabel \footnotesize \color{black}  ${19}$\fi}}] (19);

        \coordinate (25) at ([xshift=-4cm]1);
        \coordinate (27) at ([xshift=4cm]1);

\draw [color=MidnightBlue]  (22) -- (25) coordinate[c1,fill,pos=0.5,label={[label distance=-1]180:{\iflabel \footnotesize \color{black}  ${24}$\fi}}] (24) coordinate[c2,fill,label={[label distance=-1]180:{\iflabel \footnotesize \color{black}  ${25}$\fi}}] (25);
\draw [color=Mulberry] (25) -- (9) coordinate[c1,fill,pos=0.8,label={[label distance=-1]90:{\iflabel \footnotesize \color{black}  ${35}$\fi}}] (35);

\draw [color=BrickRed]  (7) -- (27) coordinate[c1,fill,pos=0.5,label={[label distance=-3]90:{\iflabel \footnotesize \color{black}  ${34}$\fi}}] (34) coordinate[c2,fill,label={[label distance=-1]90:{\iflabel \footnotesize \color{black}  ${27}$\fi}}] (27);
\draw [color=Emerald] (27) -- (23) coordinate[c1,fill,pos=0.25,label={[label distance=-1]320:{\iflabel \footnotesize \color{black}  ${26}$\fi}}] (26);

\draw [color=BlueGreen]  (10) -- (15.5,17.5) coordinate[c1,fill,pos=0.5,label={[label distance=-1]90:{\iflabel \footnotesize \color{black}  ${12}$\fi}}] (12) coordinate[c2,fill,label={[label distance=-1,xshift=5]270:{\iflabel \footnotesize \color{black}  ${13}$\fi}}] (13);
\draw [color=Tan] (13) -- (29) coordinate[c1,fill,pos=0.4,label={[label distance=-1]90:{\iflabel \footnotesize \color{black}  ${31}$\fi}}] (31);

\draw [color=RawSienna]  (28) -- (10.5,15) coordinate[c1,fill,pos=0.5,label={[label distance=-3, yshift=-3]160:{\iflabel \footnotesize \color{black}  ${30}$\fi}}] (30) coordinate[c2,fill,label={[label distance=-5]45:{\iflabel \footnotesize \color{black}  ${15}$\fi}}] (15);
\draw [color=SpringGreen] (15) -- (11) coordinate[c1,fill,pos=0.6,label={[label distance=-1]90:{\iflabel \footnotesize \color{black}  ${14}$\fi}}] (14);

\draw [color=Salmon]  (15) -- (1) coordinate[c2,fill,pos=0.2,label={[label distance=-1, yshift=2]180:{\iflabel \footnotesize \color{black}  ${17}$\fi}}] (17);
\draw [color=Fuchsia] (1)-- (13) coordinate[c2,fill,pos=0.3,label={[label distance=-1]0:{\iflabel \footnotesize \color{black}  ${16}$\fi}}] (16);

\draw [color=CornflowerBlue]  (19) -- (16) coordinate[c1,fill,pos=0.3,label={[label distance=-1]180:{\iflabel \footnotesize \color{black}  ${18}$\fi}}] (18);
\draw [color=pink] (16) -- (8) coordinate[c1,fill,pos=0.7,label={[label distance=-1]180:{\iflabel \footnotesize \color{black}  ${32}$\fi}}] (32);

\draw [color=PineGreen]  (6) -- (17) coordinate[c1,fill,pos=0.7,label={[label distance=-1, yshift=2]180:{\iflabel \footnotesize \color{black}  ${33}$\fi}}] (33);
\draw [color=DarkOrchid] (17) -- (21) coordinate[c1,fill,pos=0.4,label={[label distance=-3]20:{\iflabel \footnotesize \color{black}  ${20}$\fi}}] (20);

\draw [color=black] (25) -- (1) -- (27);

\end{tikzpicture}
\\
(c)
\end{tabular}
\caption{
\label{2018-pit-f-TIFT-TITS}
(a)
TIFS cloud, and
(b)
TITS cloud with only a single overlaid classical value assignment if the system is prepared in state $\vert 1 \rangle $~\cite{svozil-2018-whycontexts}.
(c) The combined cloud from (a) and (b) has no value assignment allowing $36=\{\}$ to be true/1;
but still allows 8 classical value assignments enumerated by Table~\ref{2018-pit-t-ACS},
with overlaid partial coverage common to all of them.
A faithful orthogonal realization is enumerated in Ref.~\cite[Table.~1, p.~102201-7]{2015-AnalyticKS}.
}
\end{figure}

For the sake of a concrete example consider the two TIFS and TITS clouds
-- that is, logics with 35 intertwined binary observables (propositions) in 24 contexts --
depicted in Fig.~\ref{2018-pit-f-TIFT-TITS}~\cite{svozil-2018-whycontexts}.
They represent quantum clouds  with the same terminal points
$ \{1 \}   \equiv  \{1' \}  $ and $  \{ 2,3,4,5,6,7 \} \equiv \{ 1',2',3',4',5' \} $,
forcing the latter ones (that is, $\{ 2,3,4,5,6,7 \} $ and $\{ 1',2',3',4',5' \} $)
to be false/0 and true/1, respectively, if the former ones (that is, $ \{1 \}   \equiv  \{1' \}  $) are true/1.

Formally, the only two-valued states on
the logics depicted in Figs.~\ref{2018-pit-f-TIFT-TITS}(a) and~\ref{2018-pit-f-TIFT-TITS}(b)
which allow $v( \{ 1 \}) = v'(\{ 1' \} )=1$
requires that $v( \{ 2,3,4,5,6,7 \} )=0$ but $v'(\{ 1',2',3',4',5' \} )=1-v( \{ 2,3,4,5,6,7 \} )$, respectively.
However,  both these logics have a faithful orthogonal representation~\cite[Table.~1, p.~102201-7]{2015-AnalyticKS} in terms of vectors
which coincide in  $\vert \{1\} \rangle =\vert \{1'\} \rangle$,
as well as in $\vert \{2,3,4,5,6,7\} \rangle = \vert \{1',2',3',4',5'\} \rangle$,
and even in all of the other adjacent observables.

The combined logic, which features 37 binary observables (propositions) in 26 contexts has no longer a classical interpretation in terms of a
partition logic, as the 8 two-valued states enumerated in Table~\ref{2018-pit-t-ACS}
cannot mutually separate~\cite[Theorem~0]{kochen1}
the observables $2$, $13$, $15$, $16$, $17$, $25$,  $27$ and $36$, respectively.
\begin{table*}
\begin{center}
\begin{ruledtabular}
\begin{tabular}{c|cccccccccccccccccccccccccccccccccccccccccccccccccccccccccccccccc}
{\#}&1&2&3&4&\multicolumn{29}{c}{\hfil$\cdots$\hfill$\cdots$\hfill$\cdots$\hfil}&34&35&36&37\\
\hline
\noalign{\vskip 0.3mm}
1&\cellcolor{red!20} 1&\cellcolor{green!20} 0& 0& 1& 0& 0& 0& 0& 0& 0& 1& 1&\cellcolor{green!20} 0& 0&\cellcolor{green!20} 0&\cellcolor{green!20} 0&\cellcolor{green!20} 0& 0& 1& 0& 1& 0& 0& 1&\cellcolor{green!20} 0& 1&\cellcolor{green!20} 0& 0& 1& 1& 0& 1& 1& 1& 1&\cellcolor{green!20} 0& 1\\
2&\cellcolor{red!20} 1&\cellcolor{green!20} 0& 0& 1& 0& 0& 0& 0& 0& 0& 1& 1&\cellcolor{green!20} 0& 0&\cellcolor{green!20} 0&\cellcolor{green!20} 0&\cellcolor{green!20} 0& 0& 1& 1& 0& 0& 1& 1&\cellcolor{green!20} 0& 0&\cellcolor{green!20} 0& 0& 0& 1& 1& 1& 1& 1& 1&\cellcolor{green!20} 0& 1\\
3&\cellcolor{red!20} 1&\cellcolor{green!20} 0& 0& 0& 1& 0& 0& 0& 0& 1& 0& 0&\cellcolor{green!20} 0& 1&\cellcolor{green!20} 0&\cellcolor{green!20} 0&\cellcolor{green!20} 0& 0& 1& 0& 1& 0& 0& 1&\cellcolor{green!20} 0& 1&\cellcolor{green!20} 0& 1& 0& 0& 1& 1& 1& 1& 1&\cellcolor{green!20} 0& 1\\
4&\cellcolor{red!20} 1&\cellcolor{green!20} 0& 0& 0& 1& 0& 0& 0& 0& 1& 0& 0&\cellcolor{green!20} 0& 1&\cellcolor{green!20} 0&\cellcolor{green!20} 0&\cellcolor{green!20} 0& 1& 0& 0& 1& 1& 0& 0&\cellcolor{green!20} 0& 1&\cellcolor{green!20} 0& 0& 0& 1& 1& 1& 1& 1& 1&\cellcolor{green!20} 0& 1\\
5&\cellcolor{red!20} 1&\cellcolor{green!20} 0& 1& 1& 0& 1& 0& 1& 0& 0& 1& 1&\cellcolor{green!20} 0& 0&\cellcolor{green!20} 0&\cellcolor{green!20} 0&\cellcolor{green!20} 0& 1& 0& 1& 0& 0& 0& 1&\cellcolor{green!20} 0& 1&\cellcolor{green!20} 0& 0& 1& 1& 0& 0& 0& 1& 1&\cellcolor{green!20} 0& 0\\
6&\cellcolor{red!20} 1&\cellcolor{green!20} 0& 1& 1& 0& 1& 0& 0& 1& 0& 0& 1&\cellcolor{green!20} 0& 1&\cellcolor{green!20} 0&\cellcolor{green!20} 0&\cellcolor{green!20} 0& 1& 0& 1& 0& 0& 0& 1&\cellcolor{green!20} 0& 1&\cellcolor{green!20} 0& 0& 1& 1& 0& 1& 0& 1& 0&\cellcolor{green!20} 0& 0\\
7&\cellcolor{red!20} 1&\cellcolor{green!20} 0& 1& 0& 1& 1& 0& 1& 0& 1& 0& 0&\cellcolor{green!20} 0& 1&\cellcolor{green!20} 0&\cellcolor{green!20} 0&\cellcolor{green!20} 0& 1& 0& 1& 0& 0& 0& 1&\cellcolor{green!20} 0& 1&\cellcolor{green!20} 0& 1& 0& 0& 1& 0& 0& 1& 1&\cellcolor{green!20} 0& 0\\
8&\cellcolor{red!20} 1&\cellcolor{green!20} 0& 1& 0& 1& 0& 1& 1& 0& 0& 0& 1&\cellcolor{green!20} 0& 1&\cellcolor{green!20} 0&\cellcolor{green!20} 0&\cellcolor{green!20} 0& 1& 0& 1& 0& 0& 0& 1&\cellcolor{green!20} 0& 1&\cellcolor{green!20} 0& 1& 0& 0& 1& 0& 1& 0& 1&\cellcolor{green!20} 0& 0\\
\end{tabular}
\end{ruledtabular}
\end{center}
\caption{Enumeration of the 8 two-valued states on 37 binary observables (propositions) of the combined quantum clouds/logics depicted in  Figs.~\ref{2018-pit-f-TIFT-TITS}(a) and~\ref{2018-pit-f-TIFT-TITS}(b).
Row vector indicate the state values on the observables, column vectors the values on all states per the respective observable.}
\label{2018-pit-t-ACS}
\end{table*}

It might be amusing to keep in mind that, because of non-separability~\cite[Theorem~0]{kochen1} of some of
the binary observables (propositions), there does not exist a proper partition logic.
However, there exist generalized urn~\cite{wright:pent,wright}  and finite automata~\cite{e-f-moore,schaller-95,schaller-96}
model realisations thereof:
just consider urns ``loaded'' with balls which have no colored symbols on them;
or no such balls at all, for the binary observables (propositions)  $2$, $13$, $15$, $16$, $17$, $25$,  $27$ and $36$.
In such cases it is no more possible to empirically reconstruct the underlying logic;
yet if an underlying logic is assumed then -- at least as long as there still are truth assignments/two-valued states on the logic -- ``reduced''
probability distributions can be defined, urns can be loaded, and automata prepared, which conform to the classical predictions from a
convex combination of these truth assignments/two-valued states --
thereby giving rise to ``reduced'' conditions of experience {\it via} hull computations.

For global/total truth assignments~\cite{pitowsky:218,hru-pit-2003}
as well as for local admissibility rules allowing partial (as opposed to total, global) truth assignments~\cite{2012-incomput-proofsCJ,2015-AnalyticKS},
such arguments can be extended to cover all terminal states which are neither collinear nor orthogonal.
One could point out that, insofar as a fixed state has to be prepared the resulting value indefiniteness/indeterminacy is state dependent.
One may indeed hold that the strongest indication for quantum value indefiniteness/indeterminacy is the {\em total absence/non-existence} of two-valued states,
as exposed in the Kochen-Specker theorem~\cite{kochen1}.
But this is rather a question of nominalistic taste, as both cases have no direct empirical testability;
and as has already been pointed out by Clifton in a private conversation in 1995: {\em ``how can you measure a contradiction?''}

\section{The ``message'' of quantum (in)determinacy}

At the peril of becoming, as expressed by Clauser~\cite{clauser-talkvie}, ``evangelical,''
let me ``sort things out'' from my own very subjective and private perspective.
(Readers adverse to ``interpretation'' and the semantic, ``meaning'' aspects of physical theory may consider stop reading at this point.)

Thereby one might be inclined to follow Planck (against Feynman~\cite{clauser-talkvie,mermin-1989-shutup,mermin-2004-shutup})
and hold it as being not too unreasonable
to take scientific comprehensibility, rationality, and causality
as a~\cite[p.~539]{Planck-32-coc}
(see also~\cite[p.~1372]{Earman20071369}) {\em ``heuristic principle, a sign-post $\ldots$ to guide us in the motley confusion of events and to show us the direction
in which scientific research must advance in order to attain fruitful results.''}

So what does all of this
--
the Born rule of quantum probabilities and its derivation by Gleason's theorem from the
Kolmogorovian axioms applied to mutually comeasurable observables,
as well as its consequences, such as
the Kochen-Specker theorem, the plethora of violations of Boole's conditions of possible experience, Pitowsky's  indeterminacy principle
and more recent extensions and variations thereof
--
``try to tell us?''

First, observe that all of the aforementioned postulates and findings are
(based upon) assumptions;  and thus consequences of the latter. Stated differently,
these findings are true not in the absolute, ontologic but in the epistemic sense:
they hold relative to the axioms or assumptions made.

Thus, in maintaining rationality
one needs to
grant oneself -- or rather one is forced to accept -- the abandonment of at least some or all assumptions made.
Some options are exotic; for instance, Itamar Pitowsky's suggestions to apply paradoxical set decompositions
to probability measures~\cite{pitowsky-83,pitowsky-86}.
Another ``exotic escape option'' is to allow only unconnected (non-intertwined) contexts whose observables are dense~\cite{godsil-zaks,meyer:99,havlicek-2000}.
Some possibilities to cope with the findings are quite straightforward,
and we shall concentrate our further attention to those~\cite{svozil-2006-omni}.

\subsection{Simultaneous definiteness of counterfactual, complementary observables, and abandonment of context independence}

Suppose one insists on the simultaneous definite omni-existence of mutually complementary, and therefore necessarily counterfactual, observables.
One straightforward way to cope with the aforementioned findings is the abandonment of context-independence of intertwining observables.

There is no indication in the quantum formalism which would support such an assumption, as the respective projection operators do not
in any way depend on the contexts involved.
However, one may hold that the outcomes are context dependent
as functions of the initial state and the context measured~\cite{svozil:040102,svozil-2011-enough,Dzhafarov-2017};
and that they actually  ``are real''
and not just ``idealistically occur in our imagination;'' that is,
being
{\em ``mental through-and-through''}~\cite{Goldschmidt2017-idealism-Ch3}.
Early conceptualizations of context-dependence aka contextuality can be found in Bohr's
remark (in his typical Nostradamus-like style)~\cite{bohr-1949}
on {\em ``the impossibility of any sharp separation
between the behavior of atomic objects and the interaction with the measuring instruments which serve to define
the conditions under which the phenomena appear.''}
Bell, referring to Bohr, suggested~\cite{bell-66}, Sec.~5) that
{\em ``the result of an observation may
reasonably depend not only on the state of the system
(including hidden variables) but also on the complete
disposition of the apparatus.''}

However, the common, prevalent, use of the term ``contextuality''
is not an explicit context-dependent form, as suggested by the realist Bell in his earlier quote,
but rather a situation where the classical predictions of quantum clouds are violated. More concretely,
if experiments on quantized systems violate certain Boole-Bell type classical bounds or direct
classical predictions,
the narratives claim to have thereby ``proven contextuality''
(e.g., see Refs.~\cite{hasegawa:230401,cabelloFilipp-2008,cabello:210401,Bartosik-09,PhysRevLett.103.160405,Bub-2010}
and Ref.~\cite{Cabello-2013-Hardylike} for a ``direct proof of quantum contextuality'').

What if we take Bell's proposal of a context dependence of valuations --
and consequently,
``classical'' contextual probability theory  -- seriously?
One of the consequences would be the introduction of an {\em uncountable multiplicity} of
counterfactual observables.
An example to illustrate this multiplicity -- comparable to de Witt's view of Everett's relative state
interpretation~\cite{everett-thesis} --  is the uncountable set of orthonormal bases of $\mathbb{R}^3$
which are all interconnected at the same single intertwining element.
A continuous angular parameter characterizes the angles between the other elements of the bases,
located in the plane orthogonal to that common intertwining element.
Contextuality suggests that the value assignment of an observable (proposition) corresponding
to this common intertwining element needs to be both true/1 and false/0, depending on the context involved,
or  whenever some quantum cloud
(collection of intertwining observables) demands
this through consistency requirements.

Indeed, the introduction of multiple quantum clouds would force any context dependence
to also implicitly depend on this general perspective --
that is, on the respective quantum cloud and its faithful orthogonal realization, which in turn determines
the quantum probabilities {\em via} the Born-Gleason rule:
Because there exist various different
quantum clouds as
``pathways interconnecting''  two observables, context dependence needs to vary according to any
concrete connection between the prepared and the measured state.

A single context participates in an arbitrary, potentially infinite, multiplicity of quantum clouds.
This requires this one context to ``behave very differently'' when it comes to contextual value assignments.
Alas, as quantum clouds are hypothetical constructions of our mind and therefore
{\em ``mental through-and-through''}~\cite{Goldschmidt2017-idealism-Ch3},
so appears context dependence: as an idealistic concept, devoid of any empirical evidence,
created to rescue the {\it desideratum} of omni-realistic existence.

Pointedly stated, contextual value assignments appear
both utterly {\it ad hoc} and abritrary -- like a {\it deus ex machina} ``saving''
the {\it desideratum} of a classical omni-value definite reality, whereby it must obey quantum probability theory
without grounding it
(indeed, in the absence of any additional criterium or principle there is no
reason to assume that the likelihood of
true/1 and false/0 is other than 50:50); as well as highly discontinuous.
In this latter, discontinuity respect, context dependence is similar to the earlier mentioned
breakup of the intertwine observables
by reducing quantum observables to disconnected contexts~\cite{godsil-zaks,meyer:99,havlicek-2000}.

It is thereby granted that these considerations apply only to cases in which the assumptions of
context independence are valid throughout the entire quantum cloud -- that is, on for every observable
in which contexts intertwine.
If this were not the case -- say, if only a single one observable
occurring in intertwining contexts is allowed to be context-dependent~\cite{svozil-2011-enough,Simmons-2017} --
the respective
clouds taylored to prove
Pitowsky's Logical Indeterminacy Principle and similar, as well as the Kochen-Specker theorems do not apply;
and therefore the aforementioned consequences are invalid.

\subsection{Abandonment of omni-value definiteness of observables in all but one context}

Nietzsche once speculated~\cite{Nietzsche-GM,Nietzsche-GOMaEH} that what he has called {\em ``slave morality''}
originated from superficially pretending that
--
in what later Blair (aka Orwell) called~\cite{Orwell-1984} {\em ``doublespeak''}
--
weakness means strength.
In a rather similar sense the lack of comprehension
-- Planck's {\em ``sign-post''} --  and even the resulting inconsistencies
tended to become reinterpreted as an asset:  nowadays
consequences of the vector-based quantum probability law are marketed as {\em ``quantum supremacy''}
--
a {\em ``quantum magic''} or {\em ``hocus-pocus''}~\cite{svozil-2016-quantum-hokus-pokus} of sorts.

Indeed, future centuries may look back at our period, and may even call it a
second ``renaissance''  period of scholasticism~\cite{specker-60}.
In years from now historians of science will be amused
about our ongoing queer efforts, the calamities and ``magic'' experienced through our
painful incapacity to recognize the obvious
--
that is, the non-existence and therefore value indefiniteness/indeterminacy of certain counterfactual observables
--
namely exactly those mentioned in Itamar Pitowsky's indeterminacy principle.

This principle has a positive interpretation
of a quantum state, defined as the maximal knowledge obtainable by simultaneous measurements
of a quantized system; or,
conversely, as the maximal information content encodable therein.
This can be formalized in terms of the
{\em value definiteness} of a single~\cite{zeil-99,svozil-2002-statepart-prl,Grangier_2002,svozil-2003-garda,svozil-2018-whycontexts} context --
or, in a more broader (non-operational) perspective, the continuum of contexts intertwined by some prepared pure quantum state
(formalized as vector or the corresponding one-dimensional orthogonal projection operator).
In terms of Hilbert space quantum mechanics this amounts to the claim that the only
value definite entity can be a single orthonormal basis/maximal operator; or a continuum of
maximal operators whose spectral sum contain proper ``true intertwines.''
All other ``observables'' grant an, albeit necessarily stochastic, value
indefinite/indeterministic, view on this
state.

If more than one context is involved we might postulate that all admissable probabilities
should at least satisfy the following criterium: they should be classical Kolmogorov-style
{\em within} any single particular context~\cite{Gleason}.
It has been suggested~\cite{Auffeves-Grangier-2017,Auffeves-Grangier-2018}
that this can be extended and formalized in a multi-context environment by a double stochastic matrix
whose entries $P( {\bf e}_i, {\bf f}_j )$, with $1 \le i,j \le n$ ($n$ is the number of distinct ``atoms'' or exclusive outcomes in each context)   are identified by the joint, conditional probabilities
of one atom ${\bf f}_j$ in the second context,
relative to a given one atom ${\bf e}_i$ in the first context.
Various types of decompositions of the double stochastic matrix correspond to various types of
probabilities:
\begin{itemize}
\item
By the Birkhoff-von Neumann theorem classical probabilities can be represented by the Birkhoff polytope
spanned by the convex hull of the set of permutation matrices:
let  $\lambda_1, \ldots , \lambda_k \ge 0$ such that $\sum_{l=1}^k\lambda_l=1$,
then
$P( {\bf e}_i, {\bf f}_j ) = \left[\sum_{l=1}^k \lambda_l \Pi_l\right]_{ij}$.
Since there exist $n!$ permutations of n elements, $k$ will be bounded from above by $k\le n!$.
Note that this type of decomposition may not be unique,
as the space spanned the permutation matrices is $\left[(n-1)^2+1\right]$-dimensional; with $n!>(n-1)^2+1$ for $n>2$.
Therefore,  the bound from above can be improved such that decompositions with $k \le (n-1)^2 +1= n^2-2(n+1)$ exist~\cite{Marcus-Ree-1959}.
Formally, a permutation matrix has a quasi-vectorial~\cite{mermin-07}
decomposition in terms of the canonical (Cartesian) basis,
such that, $\Pi_i=\sum_{j=1}^n \vert {\bf e}_j \rangle \langle {\bf e}_{\pi_i (j)} \vert $,
where $\vert {\bf e}_j \rangle$ represents the $n$-tuple
associated with the $j$th basis vector
of the canonical (Cartesian) basis, and $\pi_i(j)$ stands for the $i$th permutation of $j$.

\item
Vector based probabilities allow the following decomposition~\cite{Auffeves-Grangier-2017,Auffeves-Grangier-2018}:
$P( {\bf e}_i, {\bf f}_j ) =\text{Trace} \left(\textsf{\textbf{E}}_i \textsf{\textbf{R}}  \textsf{\textbf{F}}_j \textsf{\textbf{R}}
\right)$, where $\textsf{\textbf{E}}_i$ and $\textsf{\textbf{F}}_i$ are elements of contexts,
formalized by two sets of mutually orthogonal projection operators,
and $\textsf{\textbf{R}}$  is a real positive diagonal matrix such that the
trace of $\textsf{\textbf{R}}^2$ equals the dimension $n$, and
$\text{Trace}\left(\textsf{\textbf{E}}_i \textsf{\textbf{R}}^2
\right) = 1$. The quantum mechanical Born rule is recovered by identifying $\textsf{\textbf{R}}=\mathbb{I}_n$ with the identity matrix,
so that $P( {\bf e}_i, {\bf f}_j ) =\text{Trace} \left(\textsf{\textbf{E}}_i   \textsf{\textbf{F}}_j  \right)$.

Note that, compared to the classical case utilizing the canconical (Cartesian) basis for representing
permutations, the quantum case is less restricted in its use of bases, and, therefore,
yields an extended theta body~\cite{GroetschelLovaszSchrijver1986};
an issue the late Itamar Pitowsky became interested in~\cite{Pitowsky-08-ge}.

\item
There exist more ``exotic'' probability measures on ``reduced'' propositional spaces
such as Wright's 2-state dispersion-free measure on the pentagon/pentagram~\cite{wright:pent},
or another type of probability measure based on a discontinuous 3(2)-coloring
of the set of all unit vectors with rational coefficients~\cite{godsil-zaks,meyer:99,havlicek-2000}
whose decomposition appear to be {\it ad hoc}; at least for the time being.
\end{itemize}

Where might this aforementioned type of stochasticism arise from?
It could well be that it is introduced by interactions with the environment;
and through the many uncontrollable and, for all practical purposes~\cite{bell-a},
huge number of degrees of freedom in unknown states.

The finiteness of physical resources needs not prevent the specification of a particular vector or context.
Because any other context needs to be operationalized within the physically feasible means
available to the respective experiment: it is the measurable coordinate differences which count;
not the absolute locatedness relative to a hypothetical, idealistic absolute frame of reference
which cannot be accessed operationally.

Finally, as the type of context envisioned to be value definite can be expressed in terms of
vector spaces equipped with a scalar product
--
in particular, by identifying
a context with the corresponding
orthonormal basis
or (the spectral decomposition of) the associated maximal observable(s)
--
one may ask how one could imagine the origin of such entities?
Abstractly vectors and vector spaces could originate from a great variety of very different forms;
such as from systems of solutions of ordinary linear differential equations.
Any investigation into the origins of the quantum mechanical
Hilbert space formalism itself might,
if this turns out to be a progressive research program~\cite{lakatosch},
eventually yield to a theory indicating
operational physical capacities beyond quantum mechanics.

\section{Biographical notes on Itamar Pitowsky}

I am certainly not in the position to present a view of Itamar Pitowsky's thinking.
Therefore I shall make a few rather anecdotal observations.
First of all, he seemed to me as one of the most original physicists I have ever met
-- but that might be a triviality, given his {\it opus.}
One thing I realized was that he exhibited a -- sometimes maybe even unconscious, but sometimes very outspoken --
regret that he was working in a philosophy department.
I believe he considered himself rather a mathematician or theoretical physicist.
To this I responded that being in a philosophy department might be rather fortunate because there one could ``go wild'' in every direction;
allowing much greater freedom than in other academic realms.
But, of course, this had no effect on his uneasiness.

He was astonished that I spent a not so little money (means relative to my investment capacities)
in an Israeli internet startup company which later flopped, depriving me of all but a fraction of what I had invested.
He told me that, at least at that point,
many startups in Israel had been put up intentionally only to attract money from people like me;
only to collapse later.

A late project of his concerned quantum bounds in general;
maybe in a similar -- graph theoretical and at the time undirected to quantum -- way
as Gr{\"o}tschel, Lov{\'a}sz and Schrijver's
theta body~\cite{GroetschelLovaszSchrijver1986,Cabello-2014-gtatqc}.
The idea was not just deriving absolute~\cite{cirelson:80} or parameterized,
continuous~\cite{filipp-svo-04-qpoly,filipp-svo-04-qpoly-prl}  bounds
for existing classical conditions of possible experience obtained by hull computations of polytopes;
but rather genuine quantum bounds on, say, Einstein-Podolsky-Rosen type setups.

\begin{acknowledgments}
I kindly acknowledge enlightening criticism and suggestions by Andrew W. Simmons,
as well as discussions with Philippe Grangier on the characterization of quantum probabilities.
All remaining misconceptions and errors are mine.
\end{acknowledgments}


\begin{thebibliography}{140}%
\makeatletter
\providecommand \@ifxundefined [1]{%
 \@ifx{#1\undefined}
}%
\providecommand \@ifnum [1]{%
 \ifnum #1\expandafter \@firstoftwo
 \else \expandafter \@secondoftwo
 \fi
}%
\providecommand \@ifx [1]{%
 \ifx #1\expandafter \@firstoftwo
 \else \expandafter \@secondoftwo
 \fi
}%
\providecommand \natexlab [1]{#1}%
\providecommand \enquote  [1]{``#1''}%
\providecommand \bibnamefont  [1]{#1}%
\providecommand \bibfnamefont [1]{#1}%
\providecommand \citenamefont [1]{#1}%
\providecommand \href@noop [0]{\@secondoftwo}%
\providecommand \href [0]{\begingroup \@sanitize@url \@href}%
\providecommand \@href[1]{\@@startlink{#1}\@@href}%
\providecommand \@@href[1]{\endgroup#1\@@endlink}%
\providecommand \@sanitize@url [0]{\catcode `\\12\catcode `\$12\catcode
  `\&12\catcode `\#12\catcode `\^12\catcode `\_12\catcode `\%12\relax}%
\providecommand \@@startlink[1]{}%
\providecommand \@@endlink[0]{}%
\providecommand \url  [0]{\begingroup\@sanitize@url \@url }%
\providecommand \@url [1]{\endgroup\@href {#1}{\urlprefix }}%
\providecommand \urlprefix  [0]{URL }%
\providecommand \Eprint [0]{\href }%
\providecommand \doibase [0]{http://dx.doi.org/}%
\providecommand \selectlanguage [0]{\@gobble}%
\providecommand \bibinfo  [0]{\@secondoftwo}%
\providecommand \bibfield  [0]{\@secondoftwo}%
\providecommand \translation [1]{[#1]}%
\providecommand \BibitemOpen [0]{}%
\providecommand \bibitemStop [0]{}%
\providecommand \bibitemNoStop [0]{.\EOS\space}%
\providecommand \EOS [0]{\spacefactor3000\relax}%
\providecommand \BibitemShut  [1]{\csname bibitem#1\endcsname}%
\let\auto@bib@innerbib\@empty
\bibitem [{\citenamefont {{Rogers, Jr.}}(1967)}]{rogers1}%
  \BibitemOpen
  \bibfield  {author} {\bibinfo {author} {\bibfnamefont {Hartley}\ \bibnamefont
  {{Rogers, Jr.}}},\ }\href@noop {} {\emph {\bibinfo {title} {Theory of
  Recursive Functions and Effective Computability}}}\ (\bibinfo  {publisher}
  {MacGraw-Hill, The MIT Press},\ \bibinfo {address} {New York, Cambridge,
  MA},\ \bibinfo {year} {1967})\BibitemShut {NoStop}%
\bibitem [{\citenamefont {Odifreddi}(1989)}]{odi:89}%
  \BibitemOpen
  \bibfield  {author} {\bibinfo {author} {\bibfnamefont {Piergiorgio}\
  \bibnamefont {Odifreddi}},\ }\href@noop {} {\emph {\bibinfo {title}
  {Classical Recursion Theory, Vol. 1}}}\ (\bibinfo  {publisher}
  {North-Holland},\ \bibinfo {address} {Amsterdam},\ \bibinfo {year}
  {1989})\BibitemShut {NoStop}%
\bibitem [{\citenamefont {Smullyan}(1993)}]{Smullyan1993-SMURTF}%
  \BibitemOpen
  \bibfield  {author} {\bibinfo {author} {\bibfnamefont {Raymond~M.}\
  \bibnamefont {Smullyan}},\ }\href@noop {} {\emph {\bibinfo {title} {Recursion
  Theory for Metamathematics}}},\ Oxford Logic Guides 22\ (\bibinfo
  {publisher} {Oxford University Press},\ \bibinfo {address} {New York,
  Oxford},\ \bibinfo {year} {1993})\BibitemShut {NoStop}%
\bibitem [{\citenamefont {Kleene}(1936)}]{Kleene1936}%
  \BibitemOpen
  \bibfield  {author} {\bibinfo {author} {\bibfnamefont {Stephen~Cole}\
  \bibnamefont {Kleene}},\ }\bibfield  {title} {\enquote {\bibinfo {title}
  {General recursive functions of natural numbers},}\ }\href {\doibase
  10.1007/BF01565439} {\bibfield  {journal} {\bibinfo  {journal} {Mathematische
  Annalen}\ }\textbf {\bibinfo {volume} {112}},\ \bibinfo {pages} {727--742}
  (\bibinfo {year} {1936})}\BibitemShut {NoStop}%
\bibitem [{\citenamefont {Abbott}\ \emph {et~al.}(2012)\citenamefont {Abbott},
  \citenamefont {Calude}, \citenamefont {Conder},\ and\ \citenamefont
  {Svozil}}]{2012-incomput-proofsCJ}%
  \BibitemOpen
  \bibfield  {author} {\bibinfo {author} {\bibfnamefont {Alastair~A.}\
  \bibnamefont {Abbott}}, \bibinfo {author} {\bibfnamefont {Cristian~S.}\
  \bibnamefont {Calude}}, \bibinfo {author} {\bibfnamefont {Jonathan}\
  \bibnamefont {Conder}}, \ and\ \bibinfo {author} {\bibfnamefont {Karl}\
  \bibnamefont {Svozil}},\ }\bibfield  {title} {\enquote {\bibinfo {title}
  {Strong {K}ochen-{S}pecker theorem and incomputability of quantum
  randomness},}\ }\href {\doibase 10.1103/PhysRevA.86.062109} {\bibfield
  {journal} {\bibinfo  {journal} {Physical Review A}\ }\textbf {\bibinfo
  {volume} {86}},\ \bibinfo {pages} {062109} (\bibinfo {year} {2012})},\
  \Eprint {http://arxiv.org/abs/arXiv:1207.2029} {arXiv:1207.2029} \BibitemShut
  {NoStop}%
\bibitem [{\citenamefont {Abbott}\ \emph {et~al.}(2014)\citenamefont {Abbott},
  \citenamefont {Calude},\ and\ \citenamefont {Svozil}}]{PhysRevA.89.032109}%
  \BibitemOpen
  \bibfield  {author} {\bibinfo {author} {\bibfnamefont {Alastair~A.}\
  \bibnamefont {Abbott}}, \bibinfo {author} {\bibfnamefont {Cristian~S.}\
  \bibnamefont {Calude}}, \ and\ \bibinfo {author} {\bibfnamefont {Karl}\
  \bibnamefont {Svozil}},\ }\bibfield  {title} {\enquote {\bibinfo {title}
  {Value-indefinite observables are almost everywhere},}\ }\href {\doibase
  10.1103/PhysRevA.89.032109} {\bibfield  {journal} {\bibinfo  {journal}
  {Physical Review A}\ }\textbf {\bibinfo {volume} {89}},\ \bibinfo {pages}
  {032109} (\bibinfo {year} {2014})},\ \Eprint
  {http://arxiv.org/abs/arXiv:1309.7188} {arXiv:1309.7188} \BibitemShut
  {NoStop}%
\bibitem [{\citenamefont {Abbott}\ \emph {et~al.}(2015)\citenamefont {Abbott},
  \citenamefont {Calude},\ and\ \citenamefont {Svozil}}]{2015-AnalyticKS}%
  \BibitemOpen
  \bibfield  {author} {\bibinfo {author} {\bibfnamefont {Alastair~A.}\
  \bibnamefont {Abbott}}, \bibinfo {author} {\bibfnamefont {Cristian~S.}\
  \bibnamefont {Calude}}, \ and\ \bibinfo {author} {\bibfnamefont {Karl}\
  \bibnamefont {Svozil}},\ }\bibfield  {title} {\enquote {\bibinfo {title} {A
  variant of the {K}ochen-{S}pecker theorem localising value indefiniteness},}\
  }\href {\doibase 10.1063/1.4931658} {\bibfield  {journal} {\bibinfo
  {journal} {Journal of Mathematical Physics}\ }\textbf {\bibinfo {volume}
  {56}},\ \bibinfo {eid} {102201} (\bibinfo {year} {2015})},\ \Eprint
  {http://arxiv.org/abs/arXiv:1503.01985} {arXiv:1503.01985} \BibitemShut
  {NoStop}%
\bibitem [{\citenamefont {Pitowsky}(1989{\natexlab{a}})}]{pitowsky}%
  \BibitemOpen
  \bibfield  {author} {\bibinfo {author} {\bibfnamefont {Itamar}\ \bibnamefont
  {Pitowsky}},\ }\href {\doibase 10.1007/BFb0021186} {\emph {\bibinfo {title}
  {Quantum Probability --- Quantum Logic}}},\ \bibinfo {series} {Lecture Notes
  in Physics}, Vol.\ \bibinfo {volume} {321}\ (\bibinfo  {publisher}
  {Springer-Verlag},\ \bibinfo {address} {Berlin, Heidelberg},\ \bibinfo {year}
  {1989})\BibitemShut {NoStop}%
\bibitem [{\citenamefont {Pitowsky}(1989{\natexlab{b}})}]{pitowsky-89a}%
  \BibitemOpen
  \bibfield  {author} {\bibinfo {author} {\bibfnamefont {Itamar}\ \bibnamefont
  {Pitowsky}},\ }\bibfield  {title} {\enquote {\bibinfo {title} {From {G}eorge
  {B}oole to {J}ohn {B}ell: The origin of {B}ell's inequality},}\ }in\ \href
  {\doibase 10.1007/978-94-017-0849-4\_6} {\emph {\bibinfo {booktitle}
  {{B}ell's Theorem, Quantum Theory and the Conceptions of the Universe}}},\
  \bibinfo {series} {Fundamental Theories of Physics}, Vol.~\bibinfo {volume}
  {37},\ \bibinfo {editor} {edited by\ \bibinfo {editor} {\bibfnamefont
  {Menas}\ \bibnamefont {Kafatos}}}\ (\bibinfo  {publisher} {Kluwer Academic
  Publishers, Springer Netherlands},\ \bibinfo {address} {Dordrecht},\ \bibinfo
  {year} {1989})\ pp.\ \bibinfo {pages} {37--49}\BibitemShut {NoStop}%
\bibitem [{\citenamefont {Pitowsky}(1994)}]{Pit-94}%
  \BibitemOpen
  \bibfield  {author} {\bibinfo {author} {\bibfnamefont {Itamar}\ \bibnamefont
  {Pitowsky}},\ }\bibfield  {title} {\enquote {\bibinfo {title} {{G}eorge
  {B}oole's `conditions of possible experience' and the quantum puzzle},}\
  }\href {\doibase 10.1093/bjps/45.1.95} {\bibfield  {journal} {\bibinfo
  {journal} {The British Journal for the Philosophy of Science}\ }\textbf
  {\bibinfo {volume} {45}},\ \bibinfo {pages} {95--125} (\bibinfo {year}
  {1994})}\BibitemShut {NoStop}%
\bibitem [{\citenamefont {Pitowsky}(2006)}]{pitowsky-06}%
  \BibitemOpen
  \bibfield  {author} {\bibinfo {author} {\bibfnamefont {Itamar}\ \bibnamefont
  {Pitowsky}},\ }\bibfield  {title} {\enquote {\bibinfo {title} {Quantum
  mechanics as a theory of probability},}\ }in\ \href {\doibase
  10.1007/1-4020-4876-9_10} {\emph {\bibinfo {booktitle} {Physical Theory and
  its Interpretation}}},\ \bibinfo {series} {The Western Ontario Series in
  Philosophy of Science}, Vol.~\bibinfo {volume} {72},\ \bibinfo {editor}
  {edited by\ \bibinfo {editor} {\bibfnamefont {William}\ \bibnamefont
  {Demopoulos}}\ and\ \bibinfo {editor} {\bibfnamefont {Itamar}\ \bibnamefont
  {Pitowsky}}}\ (\bibinfo  {publisher} {Springer Netherlands},\ \bibinfo {year}
  {2006})\ pp.\ \bibinfo {pages} {213--240},\ \Eprint
  {http://arxiv.org/abs/arXiv:quant-ph/0510095} {arXiv:quant-ph/0510095}
  \BibitemShut {NoStop}%
\bibitem [{\citenamefont {Bub}\ and\ \citenamefont
  {Demopoulos}(2010)}]{BUB201085}%
  \BibitemOpen
  \bibfield  {author} {\bibinfo {author} {\bibfnamefont {Jeffrey}\ \bibnamefont
  {Bub}}\ and\ \bibinfo {author} {\bibfnamefont {Williams}\ \bibnamefont
  {Demopoulos}},\ }\bibfield  {title} {\enquote {\bibinfo {title} {{I}tamar
  {P}itowsky 1950-2010},}\ }\href {\doibase 10.1016/j.shpsb.2010.03.004}
  {\bibfield  {journal} {\bibinfo  {journal} {Studies in History and Philosophy
  of Science Part {B}: {S}tudies in History and Philosophy of Modern Physics}\
  }\textbf {\bibinfo {volume} {41}},\ \bibinfo {pages} {85} (\bibinfo {year}
  {2010})}\BibitemShut {NoStop}%
\bibitem [{\citenamefont {{ESI - The Erwin Schr\"odinger International
  Institute for Mathematical Physics}}(2001)}]{ESI-AR-2000}%
  \BibitemOpen
  \bibfield  {author} {\bibinfo {author} {\bibnamefont {{ESI - The Erwin
  Schr\"odinger International Institute for Mathematical Physics}}},\ }\href
  {https://www.esi.ac.at/material/scientific-reports-1/2000.pdf} {\enquote
  {\bibinfo {title} {Scientific report for the year 2000},}\ } (\bibinfo {year}
  {2001}),\ \bibinfo {note} {eSI-Report 2000}\BibitemShut {NoStop}%
\bibitem [{\citenamefont {Pitowsky}\ and\ \citenamefont
  {Svozil}(2001)}]{2000-poly}%
  \BibitemOpen
  \bibfield  {author} {\bibinfo {author} {\bibfnamefont {Itamar}\ \bibnamefont
  {Pitowsky}}\ and\ \bibinfo {author} {\bibfnamefont {Karl}\ \bibnamefont
  {Svozil}},\ }\bibfield  {title} {\enquote {\bibinfo {title} {New optimal
  tests of quantum nonlocality},}\ }\href {\doibase 10.1103/PhysRevA.64.014102}
  {\bibfield  {journal} {\bibinfo  {journal} {Physical Review A}\ }\textbf
  {\bibinfo {volume} {64}},\ \bibinfo {pages} {014102} (\bibinfo {year}
  {2001})},\ \Eprint {http://arxiv.org/abs/arXiv:quant-ph/0011060}
  {arXiv:quant-ph/0011060} \BibitemShut {NoStop}%
\bibitem [{\citenamefont {Sliwa}(2003)}]{sliwa-2003}%
  \BibitemOpen
  \bibfield  {author} {\bibinfo {author} {\bibfnamefont {Cezary}\ \bibnamefont
  {Sliwa}},\ }\bibfield  {title} {\enquote {\bibinfo {title} {Symmetries of the
  {B}ell correlation inequalities},}\ }\href {\doibase
  10.1016/S0375-9601(03)01115-0} {\bibfield  {journal} {\bibinfo  {journal}
  {Physics Letters A}\ }\textbf {\bibinfo {volume} {317}},\ \bibinfo {pages}
  {165--168} (\bibinfo {year} {2003})},\ \Eprint
  {http://arxiv.org/abs/arXiv:quant-ph/0305190} {arXiv:quant-ph/0305190}
  \BibitemShut {NoStop}%
\bibitem [{\citenamefont {Colins}\ and\ \citenamefont
  {Gisin}(2004)}]{collins-gisin-2003}%
  \BibitemOpen
  \bibfield  {author} {\bibinfo {author} {\bibfnamefont {Daniel}\ \bibnamefont
  {Colins}}\ and\ \bibinfo {author} {\bibfnamefont {Nicolas}\ \bibnamefont
  {Gisin}},\ }\bibfield  {title} {\enquote {\bibinfo {title} {A relevant two
  qbit {B}ell inequality inequivalent to the {CHSH} inequality},}\ }\href
  {\doibase 10.1088/0305-4470/37/5/021} {\bibfield  {journal} {\bibinfo
  {journal} {Journal of Physics A: Math. Gen.}\ }\textbf {\bibinfo {volume}
  {37}},\ \bibinfo {pages} {1775--1787} (\bibinfo {year} {2004})},\ \Eprint
  {http://arxiv.org/abs/arXiv:quant-ph/0306129} {arXiv:quant-ph/0306129}
  \BibitemShut {NoStop}%
\bibitem [{\citenamefont {Pitowsky}(1986)}]{pitowsky-86}%
  \BibitemOpen
  \bibfield  {author} {\bibinfo {author} {\bibfnamefont {Itamar}\ \bibnamefont
  {Pitowsky}},\ }\bibfield  {title} {\enquote {\bibinfo {title} {The range of
  quantum probabilities},}\ }\href@noop {} {\bibfield  {journal} {\bibinfo
  {journal} {Journal of Mathematical Physics}\ }\textbf {\bibinfo {volume}
  {27}},\ \bibinfo {pages} {1556--1565} (\bibinfo {year} {1986})}\BibitemShut
  {NoStop}%
\bibitem [{\citenamefont {Pitowsky}(1991)}]{Pit-91}%
  \BibitemOpen
  \bibfield  {author} {\bibinfo {author} {\bibfnamefont {Itamar}\ \bibnamefont
  {Pitowsky}},\ }\bibfield  {title} {\enquote {\bibinfo {title} {Correlation
  polytopes their geometry and complexity},}\ }\href {\doibase
  10.1007/BF01594946} {\bibfield  {journal} {\bibinfo  {journal} {Mathematical
  Programming}\ }\textbf {\bibinfo {volume} {50}},\ \bibinfo {pages} {395--414}
  (\bibinfo {year} {1991})}\BibitemShut {NoStop}%
\bibitem [{\citenamefont {Boole}(1862)}]{Boole-62}%
  \BibitemOpen
  \bibfield  {author} {\bibinfo {author} {\bibfnamefont {George}\ \bibnamefont
  {Boole}},\ }\bibfield  {title} {\enquote {\bibinfo {title} {On the theory of
  probabilities},}\ }\href {\doibase 10.1098/rstl.1862.0015} {\bibfield
  {journal} {\bibinfo  {journal} {Philosophical Transactions of the Royal
  Society of London}\ }\textbf {\bibinfo {volume} {152}},\ \bibinfo {pages}
  {225--252} (\bibinfo {year} {1862})}\BibitemShut {NoStop}%
\bibitem [{\citenamefont {Pitowsky}(2003)}]{Pitowsky2003395}%
  \BibitemOpen
  \bibfield  {author} {\bibinfo {author} {\bibfnamefont {Itamar}\ \bibnamefont
  {Pitowsky}},\ }\bibfield  {title} {\enquote {\bibinfo {title} {Betting on the
  outcomes of measurements: a bayesian theory of quantum probability},}\ }\href
  {\doibase 10.1016/S1355-2198(03)00035-2} {\bibfield  {journal} {\bibinfo
  {journal} {Studies in History and Philosophy of Science Part B: Studies in
  History and Philosophy of Modern Physics}\ }\textbf {\bibinfo {volume}
  {34}},\ \bibinfo {pages} {395--414} (\bibinfo {year} {2003})},\ \bibinfo
  {note} {quantum Information and Computation},\ \Eprint
  {http://arxiv.org/abs/arXiv:quant-ph/0208121} {arXiv:quant-ph/0208121}
  \BibitemShut {NoStop}%
\bibitem [{\citenamefont {Svozil}(2017{\natexlab{a}})}]{svozil-2018-c}%
  \BibitemOpen
  \bibfield  {author} {\bibinfo {author} {\bibfnamefont {Karl}\ \bibnamefont
  {Svozil}},\ }\href {https://arxiv.org/abs/1808.00813} {\enquote {\bibinfo
  {title} {Quantum clouds},}\ } (\bibinfo {year} {2017}{\natexlab{a}}),\
  \Eprint {http://arxiv.org/abs/arXiv:1808.00813} {arXiv:1808.00813}
  \BibitemShut {NoStop}%
\bibitem [{\citenamefont {Pitowsky}(1998)}]{pitowsky:218}%
  \BibitemOpen
  \bibfield  {author} {\bibinfo {author} {\bibfnamefont {Itamar}\ \bibnamefont
  {Pitowsky}},\ }\bibfield  {title} {\enquote {\bibinfo {title} {Infinite and
  finite {G}leason's theorems and the logic of indeterminacy},}\ }\href
  {\doibase 10.1063/1.532334} {\bibfield  {journal} {\bibinfo  {journal}
  {Journal of Mathematical Physics}\ }\textbf {\bibinfo {volume} {39}},\
  \bibinfo {pages} {218--228} (\bibinfo {year} {1998})}\BibitemShut {NoStop}%
\bibitem [{\citenamefont {Hrushovski}\ and\ \citenamefont
  {Pitowsky}(2004)}]{hru-pit-2003}%
  \BibitemOpen
  \bibfield  {author} {\bibinfo {author} {\bibfnamefont {Ehud}\ \bibnamefont
  {Hrushovski}}\ and\ \bibinfo {author} {\bibfnamefont {Itamar}\ \bibnamefont
  {Pitowsky}},\ }\bibfield  {title} {\enquote {\bibinfo {title}
  {Generalizations of {K}ochen and {S}pecker's theorem and the effectiveness of
  {G}leason's theorem},}\ }\href {\doibase 10.1016/j.shpsb.2003.10.002}
  {\bibfield  {journal} {\bibinfo  {journal} {Studies in History and Philosophy
  of Science Part B: Studies in History and Philosophy of Modern Physics}\
  }\textbf {\bibinfo {volume} {35}},\ \bibinfo {pages} {177--194} (\bibinfo
  {year} {2004})},\ \Eprint {http://arxiv.org/abs/arXiv:quant-ph/0307139}
  {arXiv:quant-ph/0307139} \BibitemShut {NoStop}%
\bibitem [{\citenamefont {Cabello}(2008)}]{cabello:210401}%
  \BibitemOpen
  \bibfield  {author} {\bibinfo {author} {\bibfnamefont {Ad\'an}\ \bibnamefont
  {Cabello}},\ }\bibfield  {title} {\enquote {\bibinfo {title} {Experimentally
  testable state-independent quantum contextuality},}\ }\href {\doibase
  10.1103/PhysRevLett.101.210401} {\bibfield  {journal} {\bibinfo  {journal}
  {Physical Review Letters}\ }\textbf {\bibinfo {volume} {101}},\ \bibinfo
  {eid} {210401} (\bibinfo {year} {2008})},\ \Eprint
  {http://arxiv.org/abs/arXiv:0808.2456} {arXiv:0808.2456} \BibitemShut
  {NoStop}%
\bibitem [{\citenamefont {Svozil}(2018{\natexlab{a}})}]{svozil-2016-pu-book}%
  \BibitemOpen
  \bibfield  {author} {\bibinfo {author} {\bibfnamefont {Karl}\ \bibnamefont
  {Svozil}},\ }\href {\doibase 10.1007/978-3-319-70815-7} {\emph {\bibinfo
  {title} {Physical [A]Causality. {D}eterminism, Randomness and Uncaused
  Events}}}\ (\bibinfo  {publisher} {Springer},\ \bibinfo {address} {Cham,
  Berlin, Heidelberg, New York},\ \bibinfo {year} {2018})\BibitemShut {NoStop}%
\bibitem [{\citenamefont {Svozil}(2017{\natexlab{b}})}]{svozil-2017-b}%
  \BibitemOpen
  \bibfield  {author} {\bibinfo {author} {\bibfnamefont {Karl}\ \bibnamefont
  {Svozil}},\ }\href {https://arxiv.org/abs/1707.08915} {\enquote {\bibinfo
  {title} {Classical versus quantum probabilities and correlations},}\ }
  (\bibinfo {year} {2017}{\natexlab{b}}),\ \Eprint
  {http://arxiv.org/abs/arXiv:1707.08915} {arXiv:1707.08915} \BibitemShut
  {NoStop}%
\bibitem [{\citenamefont {Bishop}\ and\ \citenamefont
  {Leeuw}(1959)}]{Bishop-Leeuw-1959}%
  \BibitemOpen
  \bibfield  {author} {\bibinfo {author} {\bibfnamefont {Errett}\ \bibnamefont
  {Bishop}}\ and\ \bibinfo {author} {\bibfnamefont {Karel~De}\ \bibnamefont
  {Leeuw}},\ }\bibfield  {title} {\enquote {\bibinfo {title} {The
  representations of linear functionals by measures on sets of extreme
  points},}\ }\href {\doibase 10.5802/aif.95} {\bibfield  {journal} {\bibinfo
  {journal} {Annals of the Fourier Institute}\ }\textbf {\bibinfo {volume}
  {9}},\ \bibinfo {pages} {305--331} (\bibinfo {year} {1959})}\BibitemShut
  {NoStop}%
\bibitem [{\citenamefont {Vorob'ev}(1962)}]{Vorobev-1962}%
  \BibitemOpen
  \bibfield  {author} {\bibinfo {author} {\bibfnamefont {N.~N.}\ \bibnamefont
  {Vorob'ev}},\ }\bibfield  {title} {\enquote {\bibinfo {title} {Consistent
  families of measures and their extensions},}\ }\href {\doibase
  10.1137/1107014} {\bibfield  {journal} {\bibinfo  {journal} {Theory of
  Probability and Its Applications}\ }\textbf {\bibinfo {volume} {7}} (\bibinfo
  {year} {1962}),\ 10.1137/1107014}\BibitemShut {NoStop}%
\bibitem [{\citenamefont {Froissart}(1981)}]{froissart-81}%
  \BibitemOpen
  \bibfield  {author} {\bibinfo {author} {\bibfnamefont {M.}~\bibnamefont
  {Froissart}},\ }\bibfield  {title} {\enquote {\bibinfo {title} {Constructive
  generalization of {B}ell's inequalities},}\ }\href
  {https://doi.org/10.1007/BF02903286} {\bibfield  {journal} {\bibinfo
  {journal} {Il Nuovo Cimento B (1971-1996)}\ }\textbf {\bibinfo {volume}
  {64}},\ \bibinfo {pages} {241--251} (\bibinfo {year} {1981})},\ \bibinfo
  {note} {10.1007/BF02903286}\BibitemShut {NoStop}%
\bibitem [{\citenamefont {{Cirel'son (=Tsirel'son)}}(1993)}]{cirelson}%
  \BibitemOpen
  \bibfield  {author} {\bibinfo {author} {\bibfnamefont {Boris~S.}\
  \bibnamefont {{Cirel'son (=Tsirel'son)}}},\ }\bibfield  {title} {\enquote
  {\bibinfo {title} {Some results and problems on quantum {B}ell-type
  inequalities},}\ }\href {http://www.tau.ac.il/~tsirel/download/hadron.pdf}
  {\bibfield  {journal} {\bibinfo  {journal} {Hadronic Journal Supplement}\
  }\textbf {\bibinfo {volume} {8}},\ \bibinfo {pages} {329--345} (\bibinfo
  {year} {1993})}\BibitemShut {NoStop}%
\bibitem [{\citenamefont {Svozil}(2001)}]{svozil-2001-cesena}%
  \BibitemOpen
  \bibfield  {author} {\bibinfo {author} {\bibfnamefont {Karl}\ \bibnamefont
  {Svozil}},\ }\href {https://arxiv.org/abs/quant-ph/0012066} {\enquote
  {\bibinfo {title} {On generalized probabilities: correlation polytopes for
  automaton logic and generalized urn models, extensions of quantum mechanics
  and parameter cheats},}\ } (\bibinfo {year} {2001}),\ \Eprint
  {http://arxiv.org/abs/arXiv:quant-ph/0012066} {arXiv:quant-ph/0012066}
  \BibitemShut {NoStop}%
\bibitem [{\citenamefont {Kochen}\ and\ \citenamefont
  {Specker}(1967)}]{kochen1}%
  \BibitemOpen
  \bibfield  {author} {\bibinfo {author} {\bibfnamefont {Simon}\ \bibnamefont
  {Kochen}}\ and\ \bibinfo {author} {\bibfnamefont {Ernst~P.}\ \bibnamefont
  {Specker}},\ }\bibfield  {title} {\enquote {\bibinfo {title} {The problem of
  hidden variables in quantum mechanics},}\ }\href {\doibase
  10.1512/iumj.1968.17.17004} {\bibfield  {journal} {\bibinfo  {journal}
  {Journal of Mathematics and Mechanics (now Indiana University Mathematics
  Journal)}\ }\textbf {\bibinfo {volume} {17}},\ \bibinfo {pages} {59--87}
  (\bibinfo {year} {1967})}\BibitemShut {NoStop}%
\bibitem [{\citenamefont {Klyachko}(2002)}]{Klyachko-2002}%
  \BibitemOpen
  \bibfield  {author} {\bibinfo {author} {\bibfnamefont {Alexander~A.}\
  \bibnamefont {Klyachko}},\ }\href {https://arxiv.org/abs/quant-ph/0206012}
  {\enquote {\bibinfo {title} {Coherent states, entanglement, and geometric
  invariant theory},}\ } (\bibinfo {year} {2002}),\ \Eprint
  {http://arxiv.org/abs/arXiv:quant-ph/0206012} {arXiv:quant-ph/0206012}
  \BibitemShut {NoStop}%
\bibitem [{\citenamefont {Klyachko}\ \emph {et~al.}(2008)\citenamefont
  {Klyachko}, \citenamefont {Can}, \citenamefont
  {Binicio\ifmmode~\breve{g}\else \u{g}\fi{}lu},\ and\ \citenamefont
  {Shumovsky}}]{Klyachko-2008}%
  \BibitemOpen
  \bibfield  {author} {\bibinfo {author} {\bibfnamefont {Alexander~A.}\
  \bibnamefont {Klyachko}}, \bibinfo {author} {\bibfnamefont {M.~Ali}\
  \bibnamefont {Can}}, \bibinfo {author} {\bibfnamefont {Sinem}\ \bibnamefont
  {Binicio\ifmmode~\breve{g}\else \u{g}\fi{}lu}}, \ and\ \bibinfo {author}
  {\bibfnamefont {Alexander~S.}\ \bibnamefont {Shumovsky}},\ }\bibfield
  {title} {\enquote {\bibinfo {title} {Simple test for hidden variables in
  spin-1 systems},}\ }\href {\doibase 10.1103/PhysRevLett.101.020403}
  {\bibfield  {journal} {\bibinfo  {journal} {Physical Review Letters}\
  }\textbf {\bibinfo {volume} {101}},\ \bibinfo {pages} {020403} (\bibinfo
  {year} {2008})},\ \Eprint {http://arxiv.org/abs/arXiv:0706.0126}
  {arXiv:0706.0126} \BibitemShut {NoStop}%
\bibitem [{\citenamefont {Bub}\ and\ \citenamefont {Stairs}(2009)}]{Bub-2009}%
  \BibitemOpen
  \bibfield  {author} {\bibinfo {author} {\bibfnamefont {Jeffrey}\ \bibnamefont
  {Bub}}\ and\ \bibinfo {author} {\bibfnamefont {Allen}\ \bibnamefont
  {Stairs}},\ }\bibfield  {title} {\enquote {\bibinfo {title} {Contextuality
  and nonlocality in `no signaling' theories},}\ }\href {\doibase
  10.1007/s10701-009-9307-8} {\bibfield  {journal} {\bibinfo  {journal}
  {Foundations of Physics}\ }\textbf {\bibinfo {volume} {39}} (\bibinfo {year}
  {2009}),\ 10.1007/s10701-009-9307-8}\BibitemShut {NoStop}%
\bibitem [{\citenamefont {Bub}\ and\ \citenamefont {Stairs}(2010)}]{Bub-2010}%
  \BibitemOpen
  \bibfield  {author} {\bibinfo {author} {\bibfnamefont {Jeffrey}\ \bibnamefont
  {Bub}}\ and\ \bibinfo {author} {\bibfnamefont {Allen}\ \bibnamefont
  {Stairs}},\ }\href {https://arxiv.org/abs/1006.0500} {\enquote {\bibinfo
  {title} {Contextuality in quantum mechanics: Testing the {K}lyachko
  inequality},}\ } (\bibinfo {year} {2010}),\ \Eprint
  {http://arxiv.org/abs/arXiv:1006.0500} {arXiv:1006.0500} \BibitemShut
  {NoStop}%
\bibitem [{\citenamefont {Badzi\c{a}g}\ \emph {et~al.}(2011)\citenamefont
  {Badzi\c{a}g}, \citenamefont {Bengtsson}, \citenamefont {Cabello},
  \citenamefont {Granstr\"om},\ and\ \citenamefont {Larsson}}]{Badziag-2011}%
  \BibitemOpen
  \bibfield  {author} {\bibinfo {author} {\bibfnamefont {Piotr}\ \bibnamefont
  {Badzi\c{a}g}}, \bibinfo {author} {\bibfnamefont {Ingemar}\ \bibnamefont
  {Bengtsson}}, \bibinfo {author} {\bibfnamefont {Ad\'an}\ \bibnamefont
  {Cabello}}, \bibinfo {author} {\bibfnamefont {Helena}\ \bibnamefont
  {Granstr\"om}}, \ and\ \bibinfo {author} {\bibfnamefont {Jan-Ake}\
  \bibnamefont {Larsson}},\ }\bibfield  {title} {\enquote {\bibinfo {title}
  {Pentagrams and paradoxes},}\ }\href {\doibase 10.1007/s10701-010-9433-3}
  {\bibfield  {journal} {\bibinfo  {journal} {Foundations of Physics}\ }\textbf
  {\bibinfo {volume} {41}} (\bibinfo {year} {2011}),\
  10.1007/s10701-010-9433-3}\BibitemShut {NoStop}%
\bibitem [{\citenamefont {Clauser}(2002)}]{clauser-talkvie}%
  \BibitemOpen
  \bibfield  {author} {\bibinfo {author} {\bibfnamefont {John}\ \bibnamefont
  {Clauser}},\ }\bibfield  {title} {\enquote {\bibinfo {title} {Early history
  of {B}ell's theorem},}\ }in\ \href {\doibase 10.1007/978-3-662-05032-3\_6}
  {\emph {\bibinfo {booktitle} {Quantum (Un)speakables: {F}rom {B}ell to
  Quantum Information}}},\ \bibinfo {editor} {edited by\ \bibinfo {editor}
  {\bibfnamefont {Reinhold}\ \bibnamefont {Bertlmann}}\ and\ \bibinfo {editor}
  {\bibfnamefont {Anton}\ \bibnamefont {Zeilinger}}}\ (\bibinfo  {publisher}
  {Springer},\ \bibinfo {address} {Berlin},\ \bibinfo {year} {2002})\ pp.\
  \bibinfo {pages} {61--96}\BibitemShut {NoStop}%
\bibitem [{\citenamefont {Stace}(1934)}]{stace}%
  \BibitemOpen
  \bibfield  {author} {\bibinfo {author} {\bibfnamefont {Walter~Terence}\
  \bibnamefont {Stace}},\ }\bibfield  {title} {\enquote {\bibinfo {title} {The
  refutation of realism},}\ }\href {\doibase 10.1093/mind/XLIII.170.145}
  {\bibfield  {journal} {\bibinfo  {journal} {Mind}\ }\textbf {\bibinfo
  {volume} {43}},\ \bibinfo {pages} {145--155} (\bibinfo {year}
  {1934})}\BibitemShut {NoStop}%
\bibitem [{\citenamefont {Peres}(1978)}]{peres222}%
  \BibitemOpen
  \bibfield  {author} {\bibinfo {author} {\bibfnamefont {Asher}\ \bibnamefont
  {Peres}},\ }\bibfield  {title} {\enquote {\bibinfo {title} {Unperformed
  experiments have no results},}\ }\href {\doibase 10.1119/1.11393} {\bibfield
  {journal} {\bibinfo  {journal} {American Journal of Physics}\ }\textbf
  {\bibinfo {volume} {46}},\ \bibinfo {pages} {745--747} (\bibinfo {year}
  {1978})}\BibitemShut {NoStop}%
\bibitem [{\citenamefont {Specker}(1960)}]{specker-60}%
  \BibitemOpen
  \bibfield  {author} {\bibinfo {author} {\bibfnamefont {Ernst}\ \bibnamefont
  {Specker}},\ }\bibfield  {title} {\enquote {\bibinfo {title} {{D}ie {L}ogik
  nicht gleichzeitig entscheidbarer {A}ussagen},}\ }\href {\doibase
  10.1111/j.1746-8361.1960.tb00422.x} {\bibfield  {journal} {\bibinfo
  {journal} {Dialectica}\ }\textbf {\bibinfo {volume} {14}},\ \bibinfo {pages}
  {239--246} (\bibinfo {year} {1960})},\ \Eprint
  {http://arxiv.org/abs/arXiv:1103.4537} {arXiv:1103.4537} \BibitemShut
  {NoStop}%
\bibitem [{\citenamefont {Specker}(2009)}]{specker-ep}%
  \BibitemOpen
  \bibfield  {author} {\bibinfo {author} {\bibfnamefont {Ernst}\ \bibnamefont
  {Specker}},\ }\href {https://vimeo.com/52923835} {\enquote {\bibinfo {title}
  {{E}rnst {S}pecker and the fundamental theorem of quantum mechanics},}\ }
  (\bibinfo {year} {2009}),\ \bibinfo {note} {video by {A}d\'an Cabello,
  recorded on June 17, 2009}\BibitemShut {NoStop}%
\bibitem [{\citenamefont {Moore}(1956)}]{e-f-moore}%
  \BibitemOpen
  \bibfield  {author} {\bibinfo {author} {\bibfnamefont {Edward~F.}\
  \bibnamefont {Moore}},\ }\bibfield  {title} {\enquote {\bibinfo {title}
  {Gedanken-experiments on sequential machines},}\ }in\ \href {\doibase
  10.1515/9781400882618-006} {\emph {\bibinfo {booktitle} {Automata Studies.
  {(AM-34)}}}},\ \bibinfo {editor} {edited by\ \bibinfo {editor} {\bibfnamefont
  {C.~E.}\ \bibnamefont {Shannon}}\ and\ \bibinfo {editor} {\bibfnamefont
  {J.}~\bibnamefont {McCarthy}}}\ (\bibinfo  {publisher} {Princeton University
  Press},\ \bibinfo {address} {Princeton, NJ},\ \bibinfo {year} {1956})\ pp.\
  \bibinfo {pages} {129--153}\BibitemShut {NoStop}%
\bibitem [{\citenamefont {Wright}(1990)}]{wright}%
  \BibitemOpen
  \bibfield  {author} {\bibinfo {author} {\bibfnamefont {Ron}\ \bibnamefont
  {Wright}},\ }\bibfield  {title} {\enquote {\bibinfo {title} {Generalized urn
  models},}\ }\href {\doibase 10.1007/BF01889696} {\bibfield  {journal}
  {\bibinfo  {journal} {Foundations of Physics}\ }\textbf {\bibinfo {volume}
  {20}},\ \bibinfo {pages} {881--903} (\bibinfo {year} {1990})}\BibitemShut
  {NoStop}%
\bibitem [{\citenamefont {Svozil}(2005)}]{svozil-2001-eua}%
  \BibitemOpen
  \bibfield  {author} {\bibinfo {author} {\bibfnamefont {Karl}\ \bibnamefont
  {Svozil}},\ }\bibfield  {title} {\enquote {\bibinfo {title} {Logical
  equivalence between generalized urn models and finite automata},}\ }\href
  {\doibase 10.1007/s10773-005-7052-0} {\bibfield  {journal} {\bibinfo
  {journal} {International Journal of Theoretical Physics}\ }\textbf {\bibinfo
  {volume} {44}},\ \bibinfo {pages} {745--754} (\bibinfo {year} {2005})},\
  \Eprint {http://arxiv.org/abs/arXiv:quant-ph/0209136}
  {arXiv:quant-ph/0209136} \BibitemShut {NoStop}%
\bibitem [{\citenamefont {Gleason}(1957)}]{Gleason}%
  \BibitemOpen
  \bibfield  {author} {\bibinfo {author} {\bibfnamefont {Andrew~M.}\
  \bibnamefont {Gleason}},\ }\bibfield  {title} {\enquote {\bibinfo {title}
  {Measures on the closed subspaces of a {H}ilbert space},}\ }\href {\doibase
  10.1512/iumj.1957.6.56050"} {\bibfield  {journal} {\bibinfo  {journal}
  {Journal of Mathematics and Mechanics (now Indiana University Mathematics
  Journal)}\ }\textbf {\bibinfo {volume} {6}},\ \bibinfo {pages} {885--893}
  (\bibinfo {year} {1957})}\BibitemShut {NoStop}%
\bibitem [{\citenamefont {Peres}(1993)}]{peres}%
  \BibitemOpen
  \bibfield  {author} {\bibinfo {author} {\bibfnamefont {Asher}\ \bibnamefont
  {Peres}},\ }\href@noop {} {\emph {\bibinfo {title} {Quantum Theory: Concepts
  and Methods}}}\ (\bibinfo  {publisher} {Kluwer Academic Publishers},\
  \bibinfo {address} {Dordrecht},\ \bibinfo {year} {1993})\BibitemShut
  {NoStop}%
\bibitem [{\citenamefont {Boole}(1854)}]{Boole}%
  \BibitemOpen
  \bibfield  {author} {\bibinfo {author} {\bibfnamefont {George}\ \bibnamefont
  {Boole}},\ }\href {http://www.gutenberg.org/ebooks/15114} {\emph {\bibinfo
  {title} {An Investigation of the Laws of Thought}}}\ (\bibinfo {year}
  {1854})\BibitemShut {NoStop}%
\bibitem [{\citenamefont {Fr\'echet}(1935)}]{Frechet1935}%
  \BibitemOpen
  \bibfield  {author} {\bibinfo {author} {\bibfnamefont {Maurice}\ \bibnamefont
  {Fr\'echet}},\ }\bibfield  {title} {\enquote {\bibinfo {title}
  {G\'en\'eralisation du th\'eor\`eme des probabilit\'es totales},}\ }\href
  {http://eudml.org/doc/212798} {\bibfield  {journal} {\bibinfo  {journal}
  {Fundamenta Mathematicae}\ }\textbf {\bibinfo {volume} {25}},\ \bibinfo
  {pages} {379--387} (\bibinfo {year} {1935})}\BibitemShut {NoStop}%
\bibitem [{\citenamefont {Hailperin}(1965)}]{Hailperin-1965}%
  \BibitemOpen
  \bibfield  {author} {\bibinfo {author} {\bibfnamefont {Theodore}\
  \bibnamefont {Hailperin}},\ }\bibfield  {title} {\enquote {\bibinfo {title}
  {Best possible inequalities for the probability of a logical function of
  events},}\ }\href {\doibase 10.2307/2313491} {\bibfield  {journal} {\bibinfo
  {journal} {The American Mathematical Monthly}\ }\textbf {\bibinfo {volume}
  {72}},\ \bibinfo {pages} {343--359} (\bibinfo {year} {1965})}\BibitemShut
  {NoStop}%
\bibitem [{\citenamefont {Hailperin}(1986)}]{Hailperin-86}%
  \BibitemOpen
  \bibfield  {author} {\bibinfo {author} {\bibfnamefont {Theodore}\
  \bibnamefont {Hailperin}},\ }\href
  {https://www.elsevier.com/books/booles-logic-and-probability/hailperin/978-0-444-87952-3}
  {\emph {\bibinfo {title} {{B}oole's Logic and Probability: {C}ritical
  Exposition from the Standpoint of Contemporary Algebra, Logic and Probability
  Theory}}},\ \bibinfo {edition} {2nd}\ ed.,\ \bibinfo {series} {Studies in
  Logic and the Foundations of Mathematics}, Vol.~\bibinfo {volume} {85}\
  (\bibinfo  {publisher} {Elsevier Science Ltd},\ \bibinfo {year}
  {1986})\BibitemShut {NoStop}%
\bibitem [{\citenamefont {Ursic}(1984)}]{Ursic1984}%
  \BibitemOpen
  \bibfield  {author} {\bibinfo {author} {\bibfnamefont {Silvio}\ \bibnamefont
  {Ursic}},\ }\bibfield  {title} {\enquote {\bibinfo {title} {A linear
  characterization of np-complete problems},}\ }in\ \href {\doibase
  10.1007/978-0-387-34768-4\_5} {\emph {\bibinfo {booktitle} {7th International
  Conference on Automated Deduction: Napa, California, USA May 14--16, 1984
  Proceedings}}},\ \bibinfo {editor} {edited by\ \bibinfo {editor}
  {\bibfnamefont {R.~E.}\ \bibnamefont {Shostak}}}\ (\bibinfo  {publisher}
  {Springer New York},\ \bibinfo {address} {New York},\ \bibinfo {year}
  {1984})\ pp.\ \bibinfo {pages} {80--100}\BibitemShut {NoStop}%
\bibitem [{\citenamefont {Ursic}(1986)}]{Ursic:1986:GFL:3023712.3023752}%
  \BibitemOpen
  \bibfield  {author} {\bibinfo {author} {\bibfnamefont {Silvio}\ \bibnamefont
  {Ursic}},\ }\bibfield  {title} {\enquote {\bibinfo {title} {Generalizing
  fuzzy logic probabilistic inferences},}\ }in\ \href
  {http://dl.acm.org/citation.cfm?id=3023712.3023752} {\emph {\bibinfo
  {booktitle} {Proceedings of the Second Conference on Uncertainty in
  Artificial Intelligence}}},\ \bibinfo {series and number} {UAI'86}\ (\bibinfo
   {publisher} {AUAI Press},\ \bibinfo {address} {Arlington, Virginia, United
  States},\ \bibinfo {year} {1986})\ pp.\ \bibinfo {pages} {303--310},\ \Eprint
  {http://arxiv.org/abs/arXiv:1304.3114} {arXiv:1304.3114} \BibitemShut
  {NoStop}%
\bibitem [{\citenamefont {Ursic}(1988)}]{Ursic1988}%
  \BibitemOpen
  \bibfield  {author} {\bibinfo {author} {\bibfnamefont {Silvio}\ \bibnamefont
  {Ursic}},\ }\bibfield  {title} {\enquote {\bibinfo {title} {Generalizing
  fuzzy logic probabilistic inferences},}\ }in\ \href@noop {} {\emph {\bibinfo
  {booktitle} {Uncertainty in Artificial Intelligence 2 {(UAI1986)}}}},\
  \bibinfo {editor} {edited by\ \bibinfo {editor} {\bibfnamefont {John~F.}\
  \bibnamefont {Lemmer}}\ and\ \bibinfo {editor} {\bibfnamefont {Laveen~N.}\
  \bibnamefont {Kanal}}}\ (\bibinfo  {publisher} {North Holland},\ \bibinfo
  {address} {Amsterdam},\ \bibinfo {year} {1988})\ pp.\ \bibinfo {pages}
  {337--362}\BibitemShut {NoStop}%
\bibitem [{\citenamefont {Beltrametti}\ and\ \citenamefont
  {Ma{\c{c}}zy{\'{n}}ski}(1991)}]{Beltrametti-1991}%
  \BibitemOpen
  \bibfield  {author} {\bibinfo {author} {\bibfnamefont {Enrico~G.}\
  \bibnamefont {Beltrametti}}\ and\ \bibinfo {author} {\bibfnamefont
  {Maciej~J.}\ \bibnamefont {Ma{\c{c}}zy{\'{n}}ski}},\ }\bibfield  {title}
  {\enquote {\bibinfo {title} {On a characterization of classical and
  nonclassical probabilities},}\ }\href {\doibase 10.1063/1.529326} {\bibfield
  {journal} {\bibinfo  {journal} {Journal of Mathematical Physics}\ }\textbf
  {\bibinfo {volume} {32}} (\bibinfo {year} {1991}),\
  10.1063/1.529326}\BibitemShut {NoStop}%
\bibitem [{\citenamefont {Pykacz}\ and\ \citenamefont
  {Santos}(1991)}]{Pykacz-1991}%
  \BibitemOpen
  \bibfield  {author} {\bibinfo {author} {\bibfnamefont {Jaros{\l}aw}\
  \bibnamefont {Pykacz}}\ and\ \bibinfo {author} {\bibfnamefont {Emilio}\
  \bibnamefont {Santos}},\ }\bibfield  {title} {\enquote {\bibinfo {title}
  {Hidden variables in quantum logic approach reexamined},}\ }\href {\doibase
  10.1063/1.529327} {\bibfield  {journal} {\bibinfo  {journal} {Journal of
  Mathematical Physics}\ }\textbf {\bibinfo {volume} {32}} (\bibinfo {year}
  {1991}),\ 10.1063/1.529327}\BibitemShut {NoStop}%
\bibitem [{\citenamefont {Sylvia}\ and\ \citenamefont
  {Majernik}(1992)}]{Pulmannova-1992}%
  \BibitemOpen
  \bibfield  {author} {\bibinfo {author} {\bibfnamefont {Pulmannov{\'{a}}}\
  \bibnamefont {Sylvia}}\ and\ \bibinfo {author} {\bibfnamefont {Vladimir}\
  \bibnamefont {Majernik}},\ }\bibfield  {title} {\enquote {\bibinfo {title}
  {{B}ell inequalities on quantum logics},}\ }\href {\doibase 10.1063/1.529638}
  {\bibfield  {journal} {\bibinfo  {journal} {Journal of Mathematical Physics}\
  }\textbf {\bibinfo {volume} {33}} (\bibinfo {year} {1992}),\
  10.1063/1.529638}\BibitemShut {NoStop}%
\bibitem [{\citenamefont {Beltrametti}\ and\ \citenamefont
  {Ma{\c{c}}zy{\'{n}}ski}(1993)}]{Beltrametti-1993}%
  \BibitemOpen
  \bibfield  {author} {\bibinfo {author} {\bibfnamefont {Enrico~G.}\
  \bibnamefont {Beltrametti}}\ and\ \bibinfo {author} {\bibfnamefont
  {Maciej~J.}\ \bibnamefont {Ma{\c{c}}zy{\'{n}}ski}},\ }\bibfield  {title}
  {\enquote {\bibinfo {title} {On the characterization of probabilities: A
  generalization of {B}ell's inequalities},}\ }\href {\doibase
  10.1063/1.530333} {\bibfield  {journal} {\bibinfo  {journal} {Journal of
  Mathematical Physics}\ }\textbf {\bibinfo {volume} {34}} (\bibinfo {year}
  {1993}),\ 10.1063/1.530333}\BibitemShut {NoStop}%
\bibitem [{\citenamefont {Beltrametti}\ and\ \citenamefont
  {Ma{\c{c}}zy{\'{n}}ski}(1994)}]{Beltrametti-1994}%
  \BibitemOpen
  \bibfield  {author} {\bibinfo {author} {\bibfnamefont {Enrico~G.}\
  \bibnamefont {Beltrametti}}\ and\ \bibinfo {author} {\bibfnamefont
  {Maciej~J.}\ \bibnamefont {Ma{\c{c}}zy{\'{n}}ski}},\ }\bibfield  {title}
  {\enquote {\bibinfo {title} {On {B}ell-type inequalities},}\ }\href {\doibase
  10.1007/bf02057861} {\bibfield  {journal} {\bibinfo  {journal} {Foundations
  of Physics}\ }\textbf {\bibinfo {volume} {24}} (\bibinfo {year} {1994}),\
  10.1007/bf02057861}\BibitemShut {NoStop}%
\bibitem [{\citenamefont {Dvure{\v{c}}enskij}\ and\ \citenamefont
  {L\"anger}(1994)}]{DvurLaen-1994}%
  \BibitemOpen
  \bibfield  {author} {\bibinfo {author} {\bibfnamefont {Anatolij}\
  \bibnamefont {Dvure{\v{c}}enskij}}\ and\ \bibinfo {author} {\bibfnamefont
  {Helmut}\ \bibnamefont {L\"anger}},\ }\bibfield  {title} {\enquote {\bibinfo
  {title} {Bell-type inequalities in horizontal sums of boolean algebras},}\
  }\href {\doibase 10.1007/bf02057864} {\bibfield  {journal} {\bibinfo
  {journal} {Foundations of Physics}\ }\textbf {\bibinfo {volume} {24}}
  (\bibinfo {year} {1994}),\ 10.1007/bf02057864}\BibitemShut {NoStop}%
\bibitem [{\citenamefont {Beltrametti}\ \emph {et~al.}(1995)\citenamefont
  {Beltrametti}, \citenamefont {{Del Noce}},\ and\ \citenamefont
  {Ma{\c{c}}zy{\'{n}}ski}}]{Beltrametti1995}%
  \BibitemOpen
  \bibfield  {author} {\bibinfo {author} {\bibfnamefont {Enrico~G.}\
  \bibnamefont {Beltrametti}}, \bibinfo {author} {\bibfnamefont {Carlo}\
  \bibnamefont {{Del Noce}}}, \ and\ \bibinfo {author} {\bibfnamefont
  {Maciej~J.}\ \bibnamefont {Ma{\c{c}}zy{\'{n}}ski}},\ }\bibfield  {title}
  {\enquote {\bibinfo {title} {Characterization and deduction of {B}ell-type
  inequalities},}\ }in\ \href {\doibase 10.1007/978-94-011-00298\_3} {\emph
  {\bibinfo {booktitle} {The Foundations of Quantum Mechanics --- Historical
  Analysis and Open Questions: Lecce, 1993}}},\ \bibinfo {editor} {edited by\
  \bibinfo {editor} {\bibfnamefont {Claudio}\ \bibnamefont {Garola}}\ and\
  \bibinfo {editor} {\bibfnamefont {Arcangelo}\ \bibnamefont {Rossi}}}\
  (\bibinfo  {publisher} {Springer Netherlands},\ \bibinfo {address}
  {Dordrecht},\ \bibinfo {year} {1995})\ pp.\ \bibinfo {pages}
  {35--41}\BibitemShut {NoStop}%
\bibitem [{\citenamefont {Beltrametti}\ and\ \citenamefont
  {Ma{\c{c}}zy{\'{n}}ski}(1995)}]{Beltrametti-1995}%
  \BibitemOpen
  \bibfield  {author} {\bibinfo {author} {\bibfnamefont {Enrico~G.}\
  \bibnamefont {Beltrametti}}\ and\ \bibinfo {author} {\bibfnamefont
  {Maciej~J.}\ \bibnamefont {Ma{\c{c}}zy{\'{n}}ski}},\ }\bibfield  {title}
  {\enquote {\bibinfo {title} {On the range of non-classical probability},}\
  }\href {\doibase 10.1016/0034-4877(96)83620-2} {\bibfield  {journal}
  {\bibinfo  {journal} {Reports on Mathematical Physics}\ }\textbf {\bibinfo
  {volume} {36}} (\bibinfo {year} {1995}),\
  10.1016/0034-4877(96)83620-2}\BibitemShut {NoStop}%
\bibitem [{\citenamefont {{Del Noce}}(1995)}]{Noce-1995}%
  \BibitemOpen
  \bibfield  {author} {\bibinfo {author} {\bibfnamefont {Carlo}\ \bibnamefont
  {{Del Noce}}},\ }\bibfield  {title} {\enquote {\bibinfo {title} {An algorithm
  for finding {B}ell-type inequalities},}\ }\href {\doibase 10.1007/bf02187346}
  {\bibfield  {journal} {\bibinfo  {journal} {Foundations of Physics Letters}\
  }\textbf {\bibinfo {volume} {8}} (\bibinfo {year} {1995}),\
  10.1007/bf02187346}\BibitemShut {NoStop}%
\bibitem [{\citenamefont {L\"anger}\ and\ \citenamefont
  {Ma{\c{c}}zy{\'{n}}ski}(1995)}]{Laenger1995}%
  \BibitemOpen
  \bibfield  {author} {\bibinfo {author} {\bibfnamefont {Helmut}\ \bibnamefont
  {L\"anger}}\ and\ \bibinfo {author} {\bibfnamefont {Maciej~J.}\ \bibnamefont
  {Ma{\c{c}}zy{\'{n}}ski}},\ }\bibfield  {title} {\enquote {\bibinfo {title}
  {On a characterization of probability measures on boolean algebras and some
  orthomodular lattices},}\ }\href {http://eudml.org/doc/32311} {\bibfield
  {journal} {\bibinfo  {journal} {Mathematica Slovaca}\ }\textbf {\bibinfo
  {volume} {45}},\ \bibinfo {pages} {455--468} (\bibinfo {year}
  {1995})}\BibitemShut {NoStop}%
\bibitem [{\citenamefont {Dvure{\v{c}}enskij}\ and\ \citenamefont
  {L\"anger}(1995{\natexlab{a}})}]{DvurLaen-1995}%
  \BibitemOpen
  \bibfield  {author} {\bibinfo {author} {\bibfnamefont {Anatolij}\
  \bibnamefont {Dvure{\v{c}}enskij}}\ and\ \bibinfo {author} {\bibfnamefont
  {Helmut}\ \bibnamefont {L\"anger}},\ }\bibfield  {title} {\enquote {\bibinfo
  {title} {{B}ell-type inequalities in orthomodular lattices. {I}.
  {I}nequalities of order 2},}\ }\href {\doibase 10.1007/bf00671363} {\bibfield
   {journal} {\bibinfo  {journal} {International Journal of Theoretical
  Physics}\ }\textbf {\bibinfo {volume} {34}} (\bibinfo {year}
  {1995}{\natexlab{a}}),\ 10.1007/bf00671363}\BibitemShut {NoStop}%
\bibitem [{\citenamefont {Dvure{\v{c}}enskij}\ and\ \citenamefont
  {L\"anger}(1995{\natexlab{b}})}]{DvurLaen-1995b}%
  \BibitemOpen
  \bibfield  {author} {\bibinfo {author} {\bibfnamefont {Anatolij}\
  \bibnamefont {Dvure{\v{c}}enskij}}\ and\ \bibinfo {author} {\bibfnamefont
  {Helmut}\ \bibnamefont {L\"anger}},\ }\bibfield  {title} {\enquote {\bibinfo
  {title} {Bell-type inequalities in orthomodular lattices. ii. inequalities of
  higher order},}\ }\href {\doibase 10.1007/bf00671364} {\bibfield  {journal}
  {\bibinfo  {journal} {International Journal of Theoretical Physics}\ }\textbf
  {\bibinfo {volume} {34}} (\bibinfo {year} {1995}{\natexlab{b}}),\
  10.1007/bf00671364}\BibitemShut {NoStop}%
\bibitem [{\citenamefont {Beltrametti}\ and\ \citenamefont
  {Bugajski}(1996)}]{Beltrametti-1996}%
  \BibitemOpen
  \bibfield  {author} {\bibinfo {author} {\bibfnamefont {E~G}\ \bibnamefont
  {Beltrametti}}\ and\ \bibinfo {author} {\bibfnamefont {S}~\bibnamefont
  {Bugajski}},\ }\bibfield  {title} {\enquote {\bibinfo {title} {The {B}ell
  phenomenon in classical frameworks},}\ }\href {\doibase
  10.1088/0305-4470/29/2/005} {\bibfield  {journal} {\bibinfo  {journal}
  {Journal of Physics A: Mathematical and General Physics}\ }\textbf {\bibinfo
  {volume} {29}} (\bibinfo {year} {1996}),\
  10.1088/0305-4470/29/2/005}\BibitemShut {NoStop}%
\bibitem [{\citenamefont {Pulmannov{\'{a}}}(2002)}]{Pulmannova-2002}%
  \BibitemOpen
  \bibfield  {author} {\bibinfo {author} {\bibfnamefont {Sylvia}\ \bibnamefont
  {Pulmannov{\'{a}}}},\ }\bibfield  {title} {\enquote {\bibinfo {title} {Hidden
  variables and {B}ell inequalities on quantum logics},}\ }\href {\doibase
  10.1023/a:1014424425657} {\bibfield  {journal} {\bibinfo  {journal}
  {Foundations of Physics}\ }\textbf {\bibinfo {volume} {32}} (\bibinfo {year}
  {2002}),\ 10.1023/a:1014424425657}\BibitemShut {NoStop}%
\bibitem [{\citenamefont {Filipp}\ and\ \citenamefont
  {Svozil}(2004{\natexlab{a}})}]{filipp-svo-04-qpoly}%
  \BibitemOpen
  \bibfield  {author} {\bibinfo {author} {\bibfnamefont {Stefan}\ \bibnamefont
  {Filipp}}\ and\ \bibinfo {author} {\bibfnamefont {Karl}\ \bibnamefont
  {Svozil}},\ }\bibfield  {title} {\enquote {\bibinfo {title} {Testing the
  bounds on quantum probabilities},}\ }\href {\doibase
  10.1103/PhysRevA.69.032101} {\bibfield  {journal} {\bibinfo  {journal}
  {Physical Review A}\ }\textbf {\bibinfo {volume} {69}},\ \bibinfo {pages}
  {032101} (\bibinfo {year} {2004}{\natexlab{a}})},\ \Eprint
  {http://arxiv.org/abs/arXiv:quant-ph/0306092} {arXiv:quant-ph/0306092}
  \BibitemShut {NoStop}%
\bibitem [{\citenamefont {Filipp}\ and\ \citenamefont
  {Svozil}(2004{\natexlab{b}})}]{filipp-svo-04-qpoly-prl}%
  \BibitemOpen
  \bibfield  {author} {\bibinfo {author} {\bibfnamefont {Stefan}\ \bibnamefont
  {Filipp}}\ and\ \bibinfo {author} {\bibfnamefont {Karl}\ \bibnamefont
  {Svozil}},\ }\bibfield  {title} {\enquote {\bibinfo {title} {Generalizing
  {T}sirelson's bound on {B}ell inequalities using a min-max principle},}\
  }\href {\doibase 10.1103/PhysRevLett.93.130407} {\bibfield  {journal}
  {\bibinfo  {journal} {Physical Review Letters}\ }\textbf {\bibinfo {volume}
  {93}},\ \bibinfo {pages} {130407} (\bibinfo {year} {2004}{\natexlab{b}})},\
  \Eprint {http://arxiv.org/abs/arXiv:quant-ph/0403175}
  {arXiv:quant-ph/0403175} \BibitemShut {NoStop}%
\bibitem [{\citenamefont {{Cirel'son (=Tsirel'son)}}(1980)}]{cirelson:80}%
  \BibitemOpen
  \bibfield  {author} {\bibinfo {author} {\bibfnamefont {Boris~S.}\
  \bibnamefont {{Cirel'son (=Tsirel'son)}}},\ }\bibfield  {title} {\enquote
  {\bibinfo {title} {Quantum generalizations of {B}ell's inequality},}\ }\href
  {\doibase 10.1007/BF00417500} {\bibfield  {journal} {\bibinfo  {journal}
  {Letters in Mathematical Physics}\ }\textbf {\bibinfo {volume} {4}},\
  \bibinfo {pages} {93--100} (\bibinfo {year} {1980})}\BibitemShut {NoStop}%
\bibitem [{\citenamefont {Weihs}\ \emph {et~al.}(1998)\citenamefont {Weihs},
  \citenamefont {Jennewein}, \citenamefont {Simon}, \citenamefont
  {Weinfurter},\ and\ \citenamefont {Zeilinger}}]{wjswz-98}%
  \BibitemOpen
  \bibfield  {author} {\bibinfo {author} {\bibfnamefont {Gregor}\ \bibnamefont
  {Weihs}}, \bibinfo {author} {\bibfnamefont {Thomas}\ \bibnamefont
  {Jennewein}}, \bibinfo {author} {\bibfnamefont {Christoph}\ \bibnamefont
  {Simon}}, \bibinfo {author} {\bibfnamefont {Harald}\ \bibnamefont
  {Weinfurter}}, \ and\ \bibinfo {author} {\bibfnamefont {Anton}\ \bibnamefont
  {Zeilinger}},\ }\bibfield  {title} {\enquote {\bibinfo {title} {Violation of
  {B}ell's inequality under strict {E}instein locality conditions},}\ }\href
  {\doibase 10.1103/PhysRevLett.81.5039} {\bibfield  {journal} {\bibinfo
  {journal} {Physical Review Letters}\ }\textbf {\bibinfo {volume} {81}},\
  \bibinfo {pages} {5039--5043} (\bibinfo {year} {1998})}\BibitemShut {NoStop}%
\bibitem [{\citenamefont {Svozil}(2009{\natexlab{a}})}]{svozil-2006-omni}%
  \BibitemOpen
  \bibfield  {author} {\bibinfo {author} {\bibfnamefont {Karl}\ \bibnamefont
  {Svozil}},\ }\bibfield  {title} {\enquote {\bibinfo {title} {Quantum
  scholasticism: On quantum contexts, counterfactuals, and the absurdities of
  quantum omniscience},}\ }\href {\doibase 10.1016/j.ins.2008.06.012}
  {\bibfield  {journal} {\bibinfo  {journal} {Information Sciences}\ }\textbf
  {\bibinfo {volume} {179}},\ \bibinfo {pages} {535--541} (\bibinfo {year}
  {2009}{\natexlab{a}})}\BibitemShut {NoStop}%
\bibitem [{\citenamefont {Cabello}\ \emph {et~al.}(2018)\citenamefont
  {Cabello}, \citenamefont {Portillo}, \citenamefont {Sol\'{i}s},\ and\
  \citenamefont {Svozil}}]{2018-minimalYIYS}%
  \BibitemOpen
  \bibfield  {author} {\bibinfo {author} {\bibfnamefont {Ad\'an}\ \bibnamefont
  {Cabello}}, \bibinfo {author} {\bibfnamefont {Jos\'{e}~R.}\ \bibnamefont
  {Portillo}}, \bibinfo {author} {\bibfnamefont {Alberto}\ \bibnamefont
  {Sol\'{i}s}}, \ and\ \bibinfo {author} {\bibfnamefont {Karl}\ \bibnamefont
  {Svozil}},\ }\bibfield  {title} {\enquote {\bibinfo {title} {Minimal
  true-implies-false and true-implies-true sets of propositions in
  noncontextual hidden-variable theories},}\ }\href {\doibase
  10.1103/PhysRevA.98.012106} {\bibfield  {journal} {\bibinfo  {journal}
  {Physical Review A}\ }\textbf {\bibinfo {volume} {98}},\ \bibinfo {pages}
  {012106} (\bibinfo {year} {2018})},\ \Eprint
  {http://arxiv.org/abs/arXiv:1805.00796} {arXiv:1805.00796} \BibitemShut
  {NoStop}%
\bibitem [{\citenamefont {Kochen}\ and\ \citenamefont
  {Specker}(1965)}]{kochen2}%
  \BibitemOpen
  \bibfield  {author} {\bibinfo {author} {\bibfnamefont {Simon}\ \bibnamefont
  {Kochen}}\ and\ \bibinfo {author} {\bibfnamefont {Ernst~P.}\ \bibnamefont
  {Specker}},\ }\bibfield  {title} {\enquote {\bibinfo {title} {Logical
  structures arising in quantum theory},}\ }in\ \href
  {https://www.elsevier.com/books/the-theory-of-models/addison/978-0-7204-2233-7}
  {\emph {\bibinfo {booktitle} {The Theory of Models, Proceedings of the 1963
  International Symposium at Berkeley}}}\ (\bibinfo  {publisher} {North
  Holland},\ \bibinfo {address} {Amsterdam, New York, Oxford},\ \bibinfo {year}
  {1965})\ pp.\ \bibinfo {pages} {177--189},\ \bibinfo {note} {reprinted in
  Ref.~\cite[pp. 209--221]{specker-ges}}\BibitemShut {NoStop}%
\bibitem [{\citenamefont {Belinfante}(1973)}]{Belinfante-73}%
  \BibitemOpen
  \bibfield  {author} {\bibinfo {author} {\bibfnamefont {Frederik~Jozef}\
  \bibnamefont {Belinfante}},\ }\href {\doibase
  10.1016/B978-0-08-017032-9.50001-7} {\emph {\bibinfo {title} {A Survey of
  Hidden-Variables Theories}}},\ \bibinfo {series} {International Series of
  Monographs in Natural Philosophy}, Vol.~\bibinfo {volume} {55}\ (\bibinfo
  {publisher} {Pergamon Press, Elsevier},\ \bibinfo {address} {Oxford, New
  York},\ \bibinfo {year} {1973})\BibitemShut {NoStop}%
\bibitem [{\citenamefont {Stairs}(1983)}]{stairs83}%
  \BibitemOpen
  \bibfield  {author} {\bibinfo {author} {\bibfnamefont {Allen}\ \bibnamefont
  {Stairs}},\ }\bibfield  {title} {\enquote {\bibinfo {title} {Quantum logic,
  realism, and value definiteness},}\ }\href {\doibase 10.1086/289140}
  {\bibfield  {journal} {\bibinfo  {journal} {Philosophy of Science}\ }\textbf
  {\bibinfo {volume} {50}},\ \bibinfo {pages} {578--602} (\bibinfo {year}
  {1983})}\BibitemShut {NoStop}%
\bibitem [{\citenamefont {Clifton}(1993)}]{clifton-93}%
  \BibitemOpen
  \bibfield  {author} {\bibinfo {author} {\bibfnamefont {Robert~K.}\
  \bibnamefont {Clifton}},\ }\bibfield  {title} {\enquote {\bibinfo {title}
  {Getting contextual and nonlocal elements-of-reality the easy way},}\ }\href
  {\doibase 10.1119/1.17239} {\bibfield  {journal} {\bibinfo  {journal}
  {American Journal of Physics}\ }\textbf {\bibinfo {volume} {61}},\ \bibinfo
  {pages} {443--447} (\bibinfo {year} {1993})}\BibitemShut {NoStop}%
\bibitem [{\citenamefont {Pt{\'{a}}k}\ and\ \citenamefont
  {Pulmannov{\'{a}}}(1991)}]{pulmannova-91}%
  \BibitemOpen
  \bibfield  {author} {\bibinfo {author} {\bibfnamefont {Pavel}\ \bibnamefont
  {Pt{\'{a}}k}}\ and\ \bibinfo {author} {\bibfnamefont {Sylvia}\ \bibnamefont
  {Pulmannov{\'{a}}}},\ }\href@noop {} {\emph {\bibinfo {title} {Orthomodular
  Structures as Quantum Logics. {I}ntrinsic Properties, State Space and
  Probabilistic Topics}}},\ \bibinfo {series} {Fundamental Theories of
  Physics}, Vol.~\bibinfo {volume} {44}\ (\bibinfo  {publisher} {Kluwer
  Academic Publishers, Springer Netherlands},\ \bibinfo {address} {Dordrecht},\
  \bibinfo {year} {1991})\BibitemShut {NoStop}%
\bibitem [{\citenamefont {Lov\'asz}\ \emph {et~al.}(1989)\citenamefont
  {Lov\'asz}, \citenamefont {Saks},\ and\ \citenamefont
  {Schrijver}}]{lovasz-89}%
  \BibitemOpen
  \bibfield  {author} {\bibinfo {author} {\bibfnamefont {L\'aszl\'o}\
  \bibnamefont {Lov\'asz}}, \bibinfo {author} {\bibfnamefont {M.}~\bibnamefont
  {Saks}}, \ and\ \bibinfo {author} {\bibfnamefont {Alexander}\ \bibnamefont
  {Schrijver}},\ }\bibfield  {title} {\enquote {\bibinfo {title} {Orthogonal
  representations and connectivity of graphs},}\ }\href {\doibase
  10.1016/0024-3795(89)90475-8} {\bibfield  {journal} {\bibinfo  {journal}
  {Linear Algebra and its Applications}\ }\textbf {\bibinfo {volume}
  {114-115}},\ \bibinfo {pages} {439--454} (\bibinfo {year} {1989})},\ \bibinfo
  {note} {special Issue Dedicated to Alan J. Hoffman}\BibitemShut {NoStop}%
\bibitem [{\citenamefont {Parsons}\ and\ \citenamefont
  {Pisanski}(1989)}]{Parsons-1989}%
  \BibitemOpen
  \bibfield  {author} {\bibinfo {author} {\bibfnamefont {T.D.}\ \bibnamefont
  {Parsons}}\ and\ \bibinfo {author} {\bibfnamefont {Tomaz}\ \bibnamefont
  {Pisanski}},\ }\bibfield  {title} {\enquote {\bibinfo {title} {Vector
  representations of graphs},}\ }\href {\doibase 10.1016/0012-365x(89)90171-4}
  {\bibfield  {journal} {\bibinfo  {journal} {Discrete Mathematics}\ }\textbf
  {\bibinfo {volume} {78}} (\bibinfo {year} {1989}),\
  10.1016/0012-365x(89)90171-4}\BibitemShut {NoStop}%
\bibitem [{\citenamefont {Cabello}\ \emph {et~al.}(2010)\citenamefont
  {Cabello}, \citenamefont {Severini},\ and\ \citenamefont
  {Winter}}]{Cabello-2010-ncoptaa}%
  \BibitemOpen
  \bibfield  {author} {\bibinfo {author} {\bibfnamefont {Ad\'an}\ \bibnamefont
  {Cabello}}, \bibinfo {author} {\bibfnamefont {Simone}\ \bibnamefont
  {Severini}}, \ and\ \bibinfo {author} {\bibfnamefont {Andreas}\ \bibnamefont
  {Winter}},\ }\href {https://arxiv.org/abs/1010.2163} {\enquote {\bibinfo
  {title} {(non-)contextuality of physical theories as an axiom},}\ } (\bibinfo
  {year} {2010}),\ \Eprint {http://arxiv.org/abs/arXiv:1010.2163}
  {arXiv:1010.2163} \BibitemShut {NoStop}%
\bibitem [{\citenamefont {Sol\'is-Encina}\ and\ \citenamefont
  {Portillo}(2015)}]{Portillo-2015}%
  \BibitemOpen
  \bibfield  {author} {\bibinfo {author} {\bibfnamefont {Alberto}\ \bibnamefont
  {Sol\'is-Encina}}\ and\ \bibinfo {author} {\bibfnamefont {Jos\'e~Ram\'on}\
  \bibnamefont {Portillo}},\ }\href {https://arxiv.org/abs/1504.03662}
  {\enquote {\bibinfo {title} {Orthogonal representation of graphs},}\ }
  (\bibinfo {year} {2015}),\ \Eprint {http://arxiv.org/abs/arXiv:1504.03662}
  {arXiv:1504.03662} \BibitemShut {NoStop}%
\bibitem [{\citenamefont {Cabello}(1994)}]{cabello-1994}%
  \BibitemOpen
  \bibfield  {author} {\bibinfo {author} {\bibfnamefont {Ad{\'{a}}n}\
  \bibnamefont {Cabello}},\ }\bibfield  {title} {\enquote {\bibinfo {title} {A
  simple proof of the {K}ochen-{S}pecker theorem},}\ }\href {\doibase
  10.1088/0143-0807/15/4/004} {\bibfield  {journal} {\bibinfo  {journal}
  {European Journal of Physics}\ }\textbf {\bibinfo {volume} {15}},\ \bibinfo
  {pages} {179--183} (\bibinfo {year} {1994})}\BibitemShut {NoStop}%
\bibitem [{\citenamefont {Cabello}(1996)}]{Cabello-1996-diss}%
  \BibitemOpen
  \bibfield  {author} {\bibinfo {author} {\bibfnamefont {Ad\'an}\ \bibnamefont
  {Cabello}},\ }\emph {\bibinfo {title} {Pruebas algebraicas de imposibilidad
  de variables ocultas en mec{\'a}nica cu{\'a}ntica}},\ \href
  {http://eprints.ucm.es/1961/1/T21049.pdf} {Ph.D. thesis},\ \bibinfo  {school}
  {Universidad Complutense de Madrid}, \bibinfo {address} {Madrid, Spain}
  (\bibinfo {year} {1996})\BibitemShut {NoStop}%
\bibitem [{\citenamefont {Johansen}(1994)}]{Johansen-1994}%
  \BibitemOpen
  \bibfield  {author} {\bibinfo {author} {\bibfnamefont {Helle~Bechmann}\
  \bibnamefont {Johansen}},\ }\bibfield  {title} {\enquote {\bibinfo {title}
  {Comment on ``getting contextual and nonlocal elements-of-reality the easy
  way''},}\ }\href {\doibase 10.1119/1.17551} {\bibfield  {journal} {\bibinfo
  {journal} {American Journal of Physics}\ }\textbf {\bibinfo {volume} {62}}
  (\bibinfo {year} {1994}),\ 10.1119/1.17551}\BibitemShut {NoStop}%
\bibitem [{\citenamefont {Vermaas}(1994)}]{Vermaas-1994}%
  \BibitemOpen
  \bibfield  {author} {\bibinfo {author} {\bibfnamefont {Pieter~E.}\
  \bibnamefont {Vermaas}},\ }\bibfield  {title} {\enquote {\bibinfo {title}
  {Comment on ``getting contextual and nonlocal elements-of-reality the easy
  way''},}\ }\href {\doibase 10.1119/1.17488} {\bibfield  {journal} {\bibinfo
  {journal} {American Journal of Physics}\ }\textbf {\bibinfo {volume} {62}}
  (\bibinfo {year} {1994}),\ 10.1119/1.17488}\BibitemShut {NoStop}%
\bibitem [{\citenamefont {Pitowsky}(1982)}]{Pitowsky-1982-subs}%
  \BibitemOpen
  \bibfield  {author} {\bibinfo {author} {\bibfnamefont {Itamar}\ \bibnamefont
  {Pitowsky}},\ }\bibfield  {title} {\enquote {\bibinfo {title} {Substitution
  and truth in quantum logic},}\ }\href {\doibase 10.2307/187281} {\bibfield
  {journal} {\bibinfo  {journal} {Philosophy of Science}\ }\textbf {\bibinfo
  {volume} {49}} (\bibinfo {year} {1982}),\ 10.2307/187281}\BibitemShut
  {NoStop}%
\bibitem [{\citenamefont {Hardy}(1992)}]{Hardy-92}%
  \BibitemOpen
  \bibfield  {author} {\bibinfo {author} {\bibfnamefont {Lucien}\ \bibnamefont
  {Hardy}},\ }\bibfield  {title} {\enquote {\bibinfo {title} {Quantum
  mechanics, local realistic theories, and lorentz-invariant realistic
  theories},}\ }\href {\doibase 10.1103/PhysRevLett.68.2981} {\bibfield
  {journal} {\bibinfo  {journal} {Physical Review Letters}\ }\textbf {\bibinfo
  {volume} {68}},\ \bibinfo {pages} {2981--2984} (\bibinfo {year}
  {1992})}\BibitemShut {NoStop}%
\bibitem [{\citenamefont {Hardy}(1993)}]{Hardy-93}%
  \BibitemOpen
  \bibfield  {author} {\bibinfo {author} {\bibfnamefont {Lucien}\ \bibnamefont
  {Hardy}},\ }\bibfield  {title} {\enquote {\bibinfo {title} {Nonlocality for
  two particles without inequalities for almost all entangled states},}\ }\href
  {\doibase 10.1103/PhysRevLett.71.1665} {\bibfield  {journal} {\bibinfo
  {journal} {Physical Review Letters}\ }\textbf {\bibinfo {volume} {71}},\
  \bibinfo {pages} {1665--1668} (\bibinfo {year} {1993})}\BibitemShut {NoStop}%
\bibitem [{\citenamefont {Boschi}\ \emph {et~al.}(1997)\citenamefont {Boschi},
  \citenamefont {Branca}, \citenamefont {De~Martini},\ and\ \citenamefont
  {Hardy}}]{hardy-97}%
  \BibitemOpen
  \bibfield  {author} {\bibinfo {author} {\bibfnamefont {D.}~\bibnamefont
  {Boschi}}, \bibinfo {author} {\bibfnamefont {S.}~\bibnamefont {Branca}},
  \bibinfo {author} {\bibfnamefont {F.}~\bibnamefont {De~Martini}}, \ and\
  \bibinfo {author} {\bibfnamefont {Lucien}\ \bibnamefont {Hardy}},\ }\bibfield
   {title} {\enquote {\bibinfo {title} {Ladder proof of nonlocality without
  inequalities: Theoretical and experimental results},}\ }\href {\doibase
  10.1103/PhysRevLett.79.2755} {\bibfield  {journal} {\bibinfo  {journal}
  {Physical Review Letters}\ }\textbf {\bibinfo {volume} {79}},\ \bibinfo
  {pages} {2755--2758} (\bibinfo {year} {1997})}\BibitemShut {NoStop}%
\bibitem [{\citenamefont {Cabello}\ and\ \citenamefont
  {Garc{\'{i}}a-Alcaine}(1995)}]{Cabello-1995-ppks}%
  \BibitemOpen
  \bibfield  {author} {\bibinfo {author} {\bibfnamefont {Ad{\'{a}}n}\
  \bibnamefont {Cabello}}\ and\ \bibinfo {author} {\bibfnamefont
  {G.}~\bibnamefont {Garc{\'{i}}a-Alcaine}},\ }\bibfield  {title} {\enquote
  {\bibinfo {title} {A hidden-variables versus quantum mechanics experiment},}\
  }\href {\doibase 10.1088/0305-4470/28/13/016} {\bibfield  {journal} {\bibinfo
   {journal} {Journal of Physics A: Mathematical and General Physics}\ }\textbf
  {\bibinfo {volume} {28}} (\bibinfo {year} {1995}),\
  10.1088/0305-4470/28/13/016}\BibitemShut {NoStop}%
\bibitem [{\citenamefont {Cabello}\ \emph {et~al.}(1996)\citenamefont
  {Cabello}, \citenamefont {Estebaranz},\ and\ \citenamefont
  {Garc{\'{i}}a-Alcaine}}]{cabello-96}%
  \BibitemOpen
  \bibfield  {author} {\bibinfo {author} {\bibfnamefont {Ad{\'{a}}n}\
  \bibnamefont {Cabello}}, \bibinfo {author} {\bibfnamefont {Jos{\'{e}}~M.}\
  \bibnamefont {Estebaranz}}, \ and\ \bibinfo {author} {\bibfnamefont
  {G.}~\bibnamefont {Garc{\'{i}}a-Alcaine}},\ }\bibfield  {title} {\enquote
  {\bibinfo {title} {{B}ell-{K}ochen-{S}pecker theorem: A proof with 18
  vectors},}\ }\href {\doibase 10.1016/0375-9601(96)00134-X} {\bibfield
  {journal} {\bibinfo  {journal} {Physics Letters A}\ }\textbf {\bibinfo
  {volume} {212}},\ \bibinfo {pages} {183--187} (\bibinfo {year} {1996})},\
  \Eprint {http://arxiv.org/abs/arXiv:quant-ph/9706009}
  {arXiv:quant-ph/9706009} \BibitemShut {NoStop}%
\bibitem [{\citenamefont {Cabello}(1997)}]{cabello-97-nhvp}%
  \BibitemOpen
  \bibfield  {author} {\bibinfo {author} {\bibfnamefont {Ad\'an}\ \bibnamefont
  {Cabello}},\ }\bibfield  {title} {\enquote {\bibinfo {title}
  {No-hidden-variables proof for two spin- particles preselected and
  postselected in unentangled states},}\ }\href {\doibase
  10.1103/PhysRevA.55.4109} {\bibfield  {journal} {\bibinfo  {journal}
  {Physical Review A}\ }\textbf {\bibinfo {volume} {55}},\ \bibinfo {pages}
  {4109--4111} (\bibinfo {year} {1997})}\BibitemShut {NoStop}%
\bibitem [{\citenamefont {Chen}\ \emph {et~al.}(2013)\citenamefont {Chen},
  \citenamefont {Cabello}, \citenamefont {Xu}, \citenamefont {Su},
  \citenamefont {Wu},\ and\ \citenamefont {Kwek}}]{Cabello-2013-HP}%
  \BibitemOpen
  \bibfield  {author} {\bibinfo {author} {\bibfnamefont {Jing-Ling}\
  \bibnamefont {Chen}}, \bibinfo {author} {\bibfnamefont {Ad\'an}\ \bibnamefont
  {Cabello}}, \bibinfo {author} {\bibfnamefont {Zhen-Peng}\ \bibnamefont {Xu}},
  \bibinfo {author} {\bibfnamefont {Hong-Yi}\ \bibnamefont {Su}}, \bibinfo
  {author} {\bibfnamefont {Chunfeng}\ \bibnamefont {Wu}}, \ and\ \bibinfo
  {author} {\bibfnamefont {L.~C.}\ \bibnamefont {Kwek}},\ }\bibfield  {title}
  {\enquote {\bibinfo {title} {Hardy's paradox for high-dimensional systems},}\
  }\href {\doibase 10.1103/PhysRevA.88.062116} {\bibfield  {journal} {\bibinfo
  {journal} {Phys. Rev. A}\ }\textbf {\bibinfo {volume} {88}},\ \bibinfo
  {pages} {062116} (\bibinfo {year} {2013})}\BibitemShut {NoStop}%
\bibitem [{\citenamefont {Cabello}\ \emph {et~al.}(2013)\citenamefont
  {Cabello}, \citenamefont {Badziag}, \citenamefont {Terra~Cunha},\ and\
  \citenamefont {Bourennane}}]{Cabello-2013-Hardylike}%
  \BibitemOpen
  \bibfield  {author} {\bibinfo {author} {\bibfnamefont {Ad\'an}\ \bibnamefont
  {Cabello}}, \bibinfo {author} {\bibfnamefont {Piotr}\ \bibnamefont
  {Badziag}}, \bibinfo {author} {\bibfnamefont {Marcelo}\ \bibnamefont
  {Terra~Cunha}}, \ and\ \bibinfo {author} {\bibfnamefont {Mohamed}\
  \bibnamefont {Bourennane}},\ }\bibfield  {title} {\enquote {\bibinfo {title}
  {Simple {H}ardy-like proof of quantum contextuality},}\ }\href {\doibase
  10.1103/PhysRevLett.111.180404} {\bibfield  {journal} {\bibinfo  {journal}
  {Physical Review Letters}\ }\textbf {\bibinfo {volume} {111}},\ \bibinfo
  {pages} {180404} (\bibinfo {year} {2013})}\BibitemShut {NoStop}%
\bibitem [{\citenamefont {Zierler}\ and\ \citenamefont
  {Schlessinger}(1965)}]{ZirlSchl-65}%
  \BibitemOpen
  \bibfield  {author} {\bibinfo {author} {\bibfnamefont {Neal}\ \bibnamefont
  {Zierler}}\ and\ \bibinfo {author} {\bibfnamefont {Michael}\ \bibnamefont
  {Schlessinger}},\ }\bibfield  {title} {\enquote {\bibinfo {title} {Boolean
  embeddings of orthomodular sets and quantum logic},}\ }\href {\doibase
  10.1215/S0012-7094-65-03224-2} {\bibfield  {journal} {\bibinfo  {journal}
  {Duke Mathematical Journal}\ }\textbf {\bibinfo {volume} {32}},\ \bibinfo
  {pages} {251--262} (\bibinfo {year} {1965})},\ \bibinfo {note} {reprinted in
  Ref.~\cite{Zierler1975}}\BibitemShut {NoStop}%
\bibitem [{\citenamefont
  {Svozil}(2018{\natexlab{b}})}]{svozil-2018-whycontexts}%
  \BibitemOpen
  \bibfield  {author} {\bibinfo {author} {\bibfnamefont {Karl}\ \bibnamefont
  {Svozil}},\ }\bibfield  {title} {\enquote {\bibinfo {title} {New forms of
  quantum value indefiniteness suggest that incompatible views on contexts are
  epistemic},}\ }\href {\doibase 10.3390/e20060406} {\bibfield  {journal}
  {\bibinfo  {journal} {Entropy}\ }\textbf {\bibinfo {volume} {20}},\ \bibinfo
  {pages} {406(22)} (\bibinfo {year} {2018}{\natexlab{b}})},\ \Eprint
  {http://arxiv.org/abs/arXiv:1804.10030} {arXiv:1804.10030} \BibitemShut
  {NoStop}%
\bibitem [{\citenamefont {Wright}(1978)}]{wright:pent}%
  \BibitemOpen
  \bibfield  {author} {\bibinfo {author} {\bibfnamefont {Ron}\ \bibnamefont
  {Wright}},\ }\bibfield  {title} {\enquote {\bibinfo {title} {The state of the
  pentagon. {A} nonclassical example},}\ }in\ \href@noop {} {\emph {\bibinfo
  {booktitle} {Mathematical Foundations of Quantum Theory}}},\ \bibinfo
  {editor} {edited by\ \bibinfo {editor} {\bibfnamefont {A.~R.}\ \bibnamefont
  {Marlow}}}\ (\bibinfo  {publisher} {Academic Press},\ \bibinfo {address} {New
  York},\ \bibinfo {year} {1978})\ pp.\ \bibinfo {pages} {255--274}\BibitemShut
  {NoStop}%
\bibitem [{\citenamefont {Schaller}\ and\ \citenamefont
  {Svozil}(1995)}]{schaller-95}%
  \BibitemOpen
  \bibfield  {author} {\bibinfo {author} {\bibfnamefont {Martin}\ \bibnamefont
  {Schaller}}\ and\ \bibinfo {author} {\bibfnamefont {Karl}\ \bibnamefont
  {Svozil}},\ }\bibfield  {title} {\enquote {\bibinfo {title} {Automaton
  partition logic versus quantum logic},}\ }\href {\doibase 10.1007/BF00676288}
  {\bibfield  {journal} {\bibinfo  {journal} {International Journal of
  Theoretical Physics}\ }\textbf {\bibinfo {volume} {34}},\ \bibinfo {pages}
  {1741--1750} (\bibinfo {year} {1995})}\BibitemShut {NoStop}%
\bibitem [{\citenamefont {Schaller}\ and\ \citenamefont
  {Svozil}(1996)}]{schaller-96}%
  \BibitemOpen
  \bibfield  {author} {\bibinfo {author} {\bibfnamefont {Martin}\ \bibnamefont
  {Schaller}}\ and\ \bibinfo {author} {\bibfnamefont {Karl}\ \bibnamefont
  {Svozil}},\ }\bibfield  {title} {\enquote {\bibinfo {title} {Automaton
  logic},}\ }\href {\doibase 10.1007/BF02302381} {\bibfield  {journal}
  {\bibinfo  {journal} {International Journal of Theoretical Physics}\ }\textbf
  {\bibinfo {volume} {35}} (\bibinfo {year} {1996}),\
  10.1007/BF02302381}\BibitemShut {NoStop}%
\bibitem [{\citenamefont {Mermin}(1989{\natexlab{a}})}]{mermin-1989-shutup}%
  \BibitemOpen
  \bibfield  {author} {\bibinfo {author} {\bibfnamefont {David~N.}\
  \bibnamefont {Mermin}},\ }\bibfield  {title} {\enquote {\bibinfo {title}
  {What's wrong with this pillow?}}\ }\href {\doibase 10.1063/1.2810963}
  {\bibfield  {journal} {\bibinfo  {journal} {Physics Today}\ }\textbf
  {\bibinfo {volume} {42}},\ \bibinfo {pages} {9--11} (\bibinfo {year}
  {1989}{\natexlab{a}})}\BibitemShut {NoStop}%
\bibitem [{\citenamefont {Mermin}(1989{\natexlab{b}})}]{mermin-2004-shutup}%
  \BibitemOpen
  \bibfield  {author} {\bibinfo {author} {\bibfnamefont {David~N.}\
  \bibnamefont {Mermin}},\ }\bibfield  {title} {\enquote {\bibinfo {title}
  {Could {F}eynman have said this?}}\ }\href {\doibase 10.1063/1.1768652}
  {\bibfield  {journal} {\bibinfo  {journal} {Physics Today}\ }\textbf
  {\bibinfo {volume} {57}},\ \bibinfo {pages} {10--11} (\bibinfo {year}
  {1989}{\natexlab{b}})}\BibitemShut {NoStop}%
\bibitem [{\citenamefont {Planck}(1932)}]{Planck-32-coc}%
  \BibitemOpen
  \bibfield  {author} {\bibinfo {author} {\bibfnamefont {Max}\ \bibnamefont
  {Planck}},\ }\bibfield  {title} {\enquote {\bibinfo {title} {The concept of
  causality},}\ }\href {https://doi.org/10.1088/0959-5309/44/5/301} {\bibfield
  {journal} {\bibinfo  {journal} {Proceedings of the Physical Society}\
  }\textbf {\bibinfo {volume} {44}},\ \bibinfo {pages} {529--539} (\bibinfo
  {year} {1932})}\BibitemShut {NoStop}%
\bibitem [{\citenamefont {Earman}(2007)}]{Earman20071369}%
  \BibitemOpen
  \bibfield  {author} {\bibinfo {author} {\bibfnamefont {John}\ \bibnamefont
  {Earman}},\ }\bibfield  {title} {\enquote {\bibinfo {title} {Aspects of
  determinism in modern physics. {P}art {B}},}\ }in\ \href {\doibase
  10.1016/B978-044451560-5/50017-8} {\emph {\bibinfo {booktitle} {Philosophy of
  Physics}}},\ \bibinfo {series and number} {Handbook of the Philosophy of
  Science},\ \bibinfo {editor} {edited by\ \bibinfo {editor} {\bibfnamefont
  {Jeremy}\ \bibnamefont {Butterfield}}\ and\ \bibinfo {editor} {\bibfnamefont
  {John}\ \bibnamefont {Earman}}}\ (\bibinfo  {publisher} {North-Holland},\
  \bibinfo {address} {Amsterdam},\ \bibinfo {year} {2007})\ pp.\ \bibinfo
  {pages} {1369--1434}\BibitemShut {NoStop}%
\bibitem [{\citenamefont {Pitowsky}(1983)}]{pitowsky-83}%
  \BibitemOpen
  \bibfield  {author} {\bibinfo {author} {\bibfnamefont {Itamar}\ \bibnamefont
  {Pitowsky}},\ }\bibfield  {title} {\enquote {\bibinfo {title} {Deterministic
  model of spin and statistics},}\ }\href {\doibase 10.1103/PhysRevD.27.2316}
  {\bibfield  {journal} {\bibinfo  {journal} {Physical Review D}\ }\textbf
  {\bibinfo {volume} {27}},\ \bibinfo {pages} {2316--2326} (\bibinfo {year}
  {1983})}\BibitemShut {NoStop}%
\bibitem [{\citenamefont {Godsil}\ and\ \citenamefont {Zaks}(1988,
  2012)}]{godsil-zaks}%
  \BibitemOpen
  \bibfield  {author} {\bibinfo {author} {\bibfnamefont {Chris~D.}\
  \bibnamefont {Godsil}}\ and\ \bibinfo {author} {\bibfnamefont
  {J.}~\bibnamefont {Zaks}},\ }\href {https://arxiv.org/abs/1201.0486}
  {\enquote {\bibinfo {title} {Coloring the sphere},}\ } (\bibinfo {year}
  {1988, 2012}),\ \bibinfo {note} {{U}niversity of {W}aterloo research report
  CORR 88-12},\ \Eprint {http://arxiv.org/abs/arXiv:1201.0486}
  {arXiv:1201.0486} \BibitemShut {NoStop}%
\bibitem [{\citenamefont {Meyer}(1999)}]{meyer:99}%
  \BibitemOpen
  \bibfield  {author} {\bibinfo {author} {\bibfnamefont {David~A.}\
  \bibnamefont {Meyer}},\ }\bibfield  {title} {\enquote {\bibinfo {title}
  {Finite precision measurement nullifies the {K}ochen-{S}pecker theorem},}\
  }\href {\doibase 10.1103/PhysRevLett.83.3751} {\bibfield  {journal} {\bibinfo
   {journal} {Physical Review Letters}\ }\textbf {\bibinfo {volume} {83}},\
  \bibinfo {pages} {3751--3754} (\bibinfo {year} {1999})},\ \Eprint
  {http://arxiv.org/abs/arXiv:quant-ph/9905080} {arXiv:quant-ph/9905080}
  \BibitemShut {NoStop}%
\bibitem [{\citenamefont {Havlicek}\ \emph {et~al.}(2001)\citenamefont
  {Havlicek}, \citenamefont {Krenn}, \citenamefont {Summhammer},\ and\
  \citenamefont {Svozil}}]{havlicek-2000}%
  \BibitemOpen
  \bibfield  {author} {\bibinfo {author} {\bibfnamefont {Hans}\ \bibnamefont
  {Havlicek}}, \bibinfo {author} {\bibfnamefont {G{\"{u}}nther}\ \bibnamefont
  {Krenn}}, \bibinfo {author} {\bibfnamefont {Johann}\ \bibnamefont
  {Summhammer}}, \ and\ \bibinfo {author} {\bibfnamefont {Karl}\ \bibnamefont
  {Svozil}},\ }\bibfield  {title} {\enquote {\bibinfo {title} {Colouring the
  rational quantum sphere and the {K}ochen-{S}pecker theorem},}\ }\href
  {\doibase 10.1088/0305-4470/34/14/312} {\bibfield  {journal} {\bibinfo
  {journal} {Journal of Physics A: Mathematical and General}\ }\textbf
  {\bibinfo {volume} {34}},\ \bibinfo {pages} {3071--3077} (\bibinfo {year}
  {2001})},\ \Eprint {http://arxiv.org/abs/arXiv:quant-ph/9911040}
  {arXiv:quant-ph/9911040} \BibitemShut {NoStop}%
\bibitem [{\citenamefont {Svozil}(2009{\natexlab{b}})}]{svozil:040102}%
  \BibitemOpen
  \bibfield  {author} {\bibinfo {author} {\bibfnamefont {Karl}\ \bibnamefont
  {Svozil}},\ }\bibfield  {title} {\enquote {\bibinfo {title} {Proposed direct
  test of a certain type of noncontextuality in quantum mechanics},}\ }\href
  {\doibase 10.1103/PhysRevA.80.040102} {\bibfield  {journal} {\bibinfo
  {journal} {Physical Review A}\ }\textbf {\bibinfo {volume} {80}},\ \bibinfo
  {eid} {040102} (\bibinfo {year} {2009}{\natexlab{b}})}\BibitemShut {NoStop}%
\bibitem [{\citenamefont {Svozil}(2012)}]{svozil-2011-enough}%
  \BibitemOpen
  \bibfield  {author} {\bibinfo {author} {\bibfnamefont {Karl}\ \bibnamefont
  {Svozil}},\ }\bibfield  {title} {\enquote {\bibinfo {title} {How much
  contextuality?}}\ }\href {\doibase 10.1007/s11047-012-9318-9} {\bibfield
  {journal} {\bibinfo  {journal} {Natural Computing}\ }\textbf {\bibinfo
  {volume} {11}},\ \bibinfo {pages} {261--265} (\bibinfo {year} {2012})},\
  \Eprint {http://arxiv.org/abs/arXiv:1103.3980} {arXiv:1103.3980} \BibitemShut
  {NoStop}%
\bibitem [{\citenamefont {Dzhafarov}\ \emph {et~al.}(2017)\citenamefont
  {Dzhafarov}, \citenamefont {Cervantes},\ and\ \citenamefont
  {Kujala}}]{Dzhafarov-2017}%
  \BibitemOpen
  \bibfield  {author} {\bibinfo {author} {\bibfnamefont {Ehtibar~N.}\
  \bibnamefont {Dzhafarov}}, \bibinfo {author} {\bibfnamefont {Victor~H.}\
  \bibnamefont {Cervantes}}, \ and\ \bibinfo {author} {\bibfnamefont
  {Janne~V.}\ \bibnamefont {Kujala}},\ }\bibfield  {title} {\enquote {\bibinfo
  {title} {Contextuality in canonical systems of random variables},}\ }\href
  {\doibase 10.1098/rsta.2016.0389} {\bibfield  {journal} {\bibinfo  {journal}
  {Philosophical Transactions of the Royal Society A: Mathematical, Physical
  and Engineering Sciences}\ }\textbf {\bibinfo {volume} {375}},\ \bibinfo
  {pages} {20160389} (\bibinfo {year} {2017})},\ \Eprint
  {http://arxiv.org/abs/arXiv:1703.01252} {arXiv:1703.01252} \BibitemShut
  {NoStop}%
\bibitem [{\citenamefont {Segal}\ and\ \citenamefont {Goldschmidt}(2017,
  2018)}]{Goldschmidt2017-idealism-Ch3}%
  \BibitemOpen
  \bibfield  {author} {\bibinfo {author} {\bibfnamefont {Aaron}\ \bibnamefont
  {Segal}}\ and\ \bibinfo {author} {\bibfnamefont {Tyron}\ \bibnamefont
  {Goldschmidt}},\ }\bibfield  {title} {\enquote {\bibinfo {title} {The
  necessity of idealism},}\ }in\ \href {\doibase
  10.1093/oso/9780198746973.003.0003} {\emph {\bibinfo {booktitle} {Idealism:
  New Essays in Metaphysics}}}\ (\bibinfo  {publisher} {Oxford University
  Press},\ \bibinfo {address} {Oxford, UK},\ \bibinfo {year} {2017, 2018})\
  pp.\ \bibinfo {pages} {34--49}\BibitemShut {NoStop}%
\bibitem [{\citenamefont {Bohr}(1949)}]{bohr-1949}%
  \BibitemOpen
  \bibfield  {author} {\bibinfo {author} {\bibfnamefont {Niels}\ \bibnamefont
  {Bohr}},\ }\bibfield  {title} {\enquote {\bibinfo {title} {Discussion with
  {E}instein on epistemological problems in atomic physics},}\ }in\ \href
  {\doibase 10.1016/S1876-0503(08)70379-7} {\emph {\bibinfo {booktitle}
  {{A}lbert {E}instein: Philosopher-Scientist}}},\ \bibinfo {editor} {edited
  by\ \bibinfo {editor} {\bibfnamefont {P.~A.}\ \bibnamefont {Schilpp}}}\
  (\bibinfo  {publisher} {The Library of Living Philosophers},\ \bibinfo
  {address} {Evanston, Ill.},\ \bibinfo {year} {1949})\ pp.\ \bibinfo {pages}
  {200--241}\BibitemShut {NoStop}%
\bibitem [{\citenamefont {Bell}(1966)}]{bell-66}%
  \BibitemOpen
  \bibfield  {author} {\bibinfo {author} {\bibfnamefont {John~Stuard}\
  \bibnamefont {Bell}},\ }\bibfield  {title} {\enquote {\bibinfo {title} {On
  the problem of hidden variables in quantum mechanics},}\ }\href {\doibase
  10.1103/RevModPhys.38.447} {\bibfield  {journal} {\bibinfo  {journal}
  {Reviews of Modern Physics}\ }\textbf {\bibinfo {volume} {38}},\ \bibinfo
  {pages} {447--452} (\bibinfo {year} {1966})}\BibitemShut {NoStop}%
\bibitem [{\citenamefont {Hasegawa}\ \emph {et~al.}(2006)\citenamefont
  {Hasegawa}, \citenamefont {Loidl}, \citenamefont {Badurek}, \citenamefont
  {Baron},\ and\ \citenamefont {Rauch}}]{hasegawa:230401}%
  \BibitemOpen
  \bibfield  {author} {\bibinfo {author} {\bibfnamefont {Yuji}\ \bibnamefont
  {Hasegawa}}, \bibinfo {author} {\bibfnamefont {Rudolf}\ \bibnamefont
  {Loidl}}, \bibinfo {author} {\bibfnamefont {Gerald}\ \bibnamefont {Badurek}},
  \bibinfo {author} {\bibfnamefont {Matthias}\ \bibnamefont {Baron}}, \ and\
  \bibinfo {author} {\bibfnamefont {Helmut}\ \bibnamefont {Rauch}},\ }\bibfield
   {title} {\enquote {\bibinfo {title} {Quantum contextuality in a
  single-neutron optical experiment},}\ }\href {\doibase
  10.1103/PhysRevLett.97.230401} {\bibfield  {journal} {\bibinfo  {journal}
  {Physical Review Letters}\ }\textbf {\bibinfo {volume} {97}},\ \bibinfo {eid}
  {230401} (\bibinfo {year} {2006})}\BibitemShut {NoStop}%
\bibitem [{\citenamefont {Cabello}\ \emph {et~al.}(2008)\citenamefont
  {Cabello}, \citenamefont {Filipp}, \citenamefont {Rauch},\ and\ \citenamefont
  {Hasegawa}}]{cabelloFilipp-2008}%
  \BibitemOpen
  \bibfield  {author} {\bibinfo {author} {\bibfnamefont {Ad\'an}\ \bibnamefont
  {Cabello}}, \bibinfo {author} {\bibfnamefont {Stefan}\ \bibnamefont
  {Filipp}}, \bibinfo {author} {\bibfnamefont {Helmut}\ \bibnamefont {Rauch}},
  \ and\ \bibinfo {author} {\bibfnamefont {Yuji}\ \bibnamefont {Hasegawa}},\
  }\bibfield  {title} {\enquote {\bibinfo {title} {Proposed experiment for
  testing quantum contextuality with neutrons},}\ }\href {\doibase
  10.1103/PhysRevLett.100.130404} {\bibfield  {journal} {\bibinfo  {journal}
  {Physical Review Letters}\ }\textbf {\bibinfo {volume} {100}},\ \bibinfo
  {pages} {130404} (\bibinfo {year} {2008})}\BibitemShut {NoStop}%
\bibitem [{\citenamefont {Bartosik}\ \emph {et~al.}(2009)\citenamefont
  {Bartosik}, \citenamefont {Klepp}, \citenamefont {Schmitzer}, \citenamefont
  {Sponar}, \citenamefont {Cabello}, \citenamefont {Rauch},\ and\ \citenamefont
  {Hasegawa}}]{Bartosik-09}%
  \BibitemOpen
  \bibfield  {author} {\bibinfo {author} {\bibfnamefont {H.}~\bibnamefont
  {Bartosik}}, \bibinfo {author} {\bibfnamefont {J.}~\bibnamefont {Klepp}},
  \bibinfo {author} {\bibfnamefont {C.}~\bibnamefont {Schmitzer}}, \bibinfo
  {author} {\bibfnamefont {S.}~\bibnamefont {Sponar}}, \bibinfo {author}
  {\bibfnamefont {A.}~\bibnamefont {Cabello}}, \bibinfo {author} {\bibfnamefont
  {H.}~\bibnamefont {Rauch}}, \ and\ \bibinfo {author} {\bibfnamefont
  {Y.}~\bibnamefont {Hasegawa}},\ }\bibfield  {title} {\enquote {\bibinfo
  {title} {Experimental test of quantum contextuality in neutron
  interferometry},}\ }\href {\doibase 10.1103/PhysRevLett.103.040403}
  {\bibfield  {journal} {\bibinfo  {journal} {Physical Review Letters}\
  }\textbf {\bibinfo {volume} {103}},\ \bibinfo {pages} {040403} (\bibinfo
  {year} {2009})},\ \Eprint {http://arxiv.org/abs/arXiv:0904.4576}
  {arXiv:0904.4576} \BibitemShut {NoStop}%
\bibitem [{\citenamefont {Amselem}\ \emph {et~al.}(2009)\citenamefont
  {Amselem}, \citenamefont {R\aa{}dmark}, \citenamefont {Bourennane},\ and\
  \citenamefont {Cabello}}]{PhysRevLett.103.160405}%
  \BibitemOpen
  \bibfield  {author} {\bibinfo {author} {\bibfnamefont {Elias}\ \bibnamefont
  {Amselem}}, \bibinfo {author} {\bibfnamefont {Magnus}\ \bibnamefont
  {R\aa{}dmark}}, \bibinfo {author} {\bibfnamefont {Mohamed}\ \bibnamefont
  {Bourennane}}, \ and\ \bibinfo {author} {\bibfnamefont {Ad\'an}\ \bibnamefont
  {Cabello}},\ }\bibfield  {title} {\enquote {\bibinfo {title}
  {State-independent quantum contextuality with single photons},}\ }\href
  {\doibase 10.1103/PhysRevLett.103.160405} {\bibfield  {journal} {\bibinfo
  {journal} {Physical Review Letters}\ }\textbf {\bibinfo {volume} {103}},\
  \bibinfo {pages} {160405} (\bibinfo {year} {2009})}\BibitemShut {NoStop}%
\bibitem [{\citenamefont {{Everett III}}(1973)}]{everett-thesis}%
  \BibitemOpen
  \bibfield  {author} {\bibinfo {author} {\bibfnamefont {Hugh}\ \bibnamefont
  {{Everett III}}},\ }\bibfield  {title} {\enquote {\bibinfo {title} {The
  many-worlds interpretation of quantum mechanics},}\ \ }(\bibinfo  {publisher}
  {Princeton University Press},\ \bibinfo {address} {Princeton, NJ},\ \bibinfo
  {year} {1973})\ pp.\ \bibinfo {pages} {3--140}\BibitemShut {NoStop}%
\bibitem [{\citenamefont {Simmons}(2017)}]{Simmons-2017}%
  \BibitemOpen
  \bibfield  {author} {\bibinfo {author} {\bibfnamefont {Andrew~W.}\
  \bibnamefont {Simmons}},\ }\href {https://arxiv.org/abs/1712.03766} {\enquote
  {\bibinfo {title} {How (maximally) contextual is quantum mechanics?}}\ }
  (\bibinfo {year} {2017}),\ \Eprint {http://arxiv.org/abs/arXiv:1712.03766}
  {arXiv:1712.03766} \BibitemShut {NoStop}%
\bibitem [{\citenamefont {Nietzsche}(1887, 2009-)}]{Nietzsche-GM}%
  \BibitemOpen
  \bibfield  {author} {\bibinfo {author} {\bibfnamefont {Friedrich}\
  \bibnamefont {Nietzsche}},\ }\href
  {http://www.nietzschesource.org/\#eKGWB/GM} {\emph {\bibinfo {title} {{Z}ur
  {G}enealogie der {M}oral (On the Genealogy of Morality)}}}\ (\bibinfo {year}
  {1887, 2009-})\ \bibinfo {note} {digital critical edition of the complete
  works and letters, based on the critical text by G. Colli and M. Montinari,
  Berlin/New York, de Gruyter 1967-, edited by Paolo D'Iorio}\BibitemShut
  {NoStop}%
\bibitem [{\citenamefont {Nietzsche}(1887, 1908; 1989,
  2010)}]{Nietzsche-GOMaEH}%
  \BibitemOpen
  \bibfield  {author} {\bibinfo {author} {\bibfnamefont {Friedrich~Wilhelm}\
  \bibnamefont {Nietzsche}},\ }\href
  {https://www.penguinrandomhouse.com/books/121939/on-the-genealogy-of-morals-and-ecce-homo-by-friedrich-nietzsche-edited-with-a-commentary-by-walter-kaufmann/9780679724629/}
  {\emph {\bibinfo {title} {On the Genealogy of Morals and Ecce Homo}}}\
  (\bibinfo  {publisher} {Vintage, Penguin, Random House},\ \bibinfo {year}
  {1887, 1908; 1989, 2010})\ \bibinfo {note} {translated by Walter Arnold
  Kaufmann}\BibitemShut {NoStop}%
\bibitem [{\citenamefont {(aka George~Orwell)}(1949)}]{Orwell-1984}%
  \BibitemOpen
  \bibfield  {author} {\bibinfo {author} {\bibfnamefont {Eric Arthur~Blair}\
  \bibnamefont {(aka George~Orwell)}},\ }\href
  {http://gutenberg.net.au/ebooks01/0100021.txt} {\emph {\bibinfo {title}
  {Nineteen Eighty-Four (aka 1984)}}},\ Twentieth century classics\ (\bibinfo
  {publisher} {Secker \& Warburg},\ \bibinfo {address} {Cambridge, MA},\
  \bibinfo {year} {1949})\BibitemShut {NoStop}%
\bibitem [{\citenamefont {Svozil}(2016)}]{svozil-2016-quantum-hokus-pokus}%
  \BibitemOpen
  \bibfield  {author} {\bibinfo {author} {\bibfnamefont {Karl}\ \bibnamefont
  {Svozil}},\ }\bibfield  {title} {\enquote {\bibinfo {title} {Quantum
  hocus-pocus},}\ }\href {\doibase 10.3354/esep00171} {\bibfield  {journal}
  {\bibinfo  {journal} {Ethics in Science and Environmental Politics (ESEP)}\
  }\textbf {\bibinfo {volume} {16}},\ \bibinfo {pages} {25--30} (\bibinfo
  {year} {2016})},\ \Eprint {http://arxiv.org/abs/arXiv:1605.08569}
  {arXiv:1605.08569} \BibitemShut {NoStop}%
\bibitem [{\citenamefont {Zeilinger}(1999)}]{zeil-99}%
  \BibitemOpen
  \bibfield  {author} {\bibinfo {author} {\bibfnamefont {Anton}\ \bibnamefont
  {Zeilinger}},\ }\bibfield  {title} {\enquote {\bibinfo {title} {A
  foundational principle for quantum mechanics},}\ }\href {\doibase
  10.1023/A:1018820410908} {\bibfield  {journal} {\bibinfo  {journal}
  {Foundations of Physics}\ }\textbf {\bibinfo {volume} {29}},\ \bibinfo
  {pages} {631--643} (\bibinfo {year} {1999})}\BibitemShut {NoStop}%
\bibitem [{\citenamefont {Svozil}(2002)}]{svozil-2002-statepart-prl}%
  \BibitemOpen
  \bibfield  {author} {\bibinfo {author} {\bibfnamefont {Karl}\ \bibnamefont
  {Svozil}},\ }\bibfield  {title} {\enquote {\bibinfo {title} {Quantum
  information in base $n$ defined by state partitions},}\ }\href {\doibase
  10.1103/PhysRevA.66.044306} {\bibfield  {journal} {\bibinfo  {journal}
  {Physical Review A}\ }\textbf {\bibinfo {volume} {66}},\ \bibinfo {pages}
  {044306} (\bibinfo {year} {2002})},\ \Eprint
  {http://arxiv.org/abs/arXiv:quant-ph/0205031} {arXiv:quant-ph/0205031}
  \BibitemShut {NoStop}%
\bibitem [{\citenamefont {Grangier}(2002)}]{Grangier_2002}%
  \BibitemOpen
  \bibfield  {author} {\bibinfo {author} {\bibfnamefont {Philippe}\
  \bibnamefont {Grangier}},\ }\bibfield  {title} {\enquote {\bibinfo {title}
  {Contextual objectivity: a realistic interpretation of quantum mechanics},}\
  }\href {\doibase 10.1088/0143-0807/23/3/312} {\bibfield  {journal} {\bibinfo
  {journal} {European Journal of Physics}\ }\textbf {\bibinfo {volume} {23}},\
  \bibinfo {pages} {331--337} (\bibinfo {year} {2002})},\ \Eprint
  {http://arxiv.org/abs/arXiv:quant-ph/0012122} {arXiv:quant-ph/0012122}
  \BibitemShut {NoStop}%
\bibitem [{\citenamefont {Svozil}(2004)}]{svozil-2003-garda}%
  \BibitemOpen
  \bibfield  {author} {\bibinfo {author} {\bibfnamefont {Karl}\ \bibnamefont
  {Svozil}},\ }\bibfield  {title} {\enquote {\bibinfo {title} {Quantum
  information via state partitions and the context translation principle},}\
  }\href {\doibase 10.1080/09500340410001664179} {\bibfield  {journal}
  {\bibinfo  {journal} {Journal of Modern Optics}\ }\textbf {\bibinfo {volume}
  {51}},\ \bibinfo {pages} {811--819} (\bibinfo {year} {2004})},\ \Eprint
  {http://arxiv.org/abs/arXiv:quant-ph/0308110} {arXiv:quant-ph/0308110}
  \BibitemShut {NoStop}%
\bibitem [{\citenamefont {Auff\'eves}\ and\ \citenamefont
  {Grangier}(2017)}]{Auffeves-Grangier-2017}%
  \BibitemOpen
  \bibfield  {author} {\bibinfo {author} {\bibfnamefont {Alexia}\ \bibnamefont
  {Auff\'eves}}\ and\ \bibinfo {author} {\bibfnamefont {Philippe}\ \bibnamefont
  {Grangier}},\ }\bibfield  {title} {\enquote {\bibinfo {title} {Recovering the
  quantum formalism from physically realist axioms},}\ }\href {\doibase
  10.1038/srep43365} {\bibfield  {journal} {\bibinfo  {journal} {Scientific
  Reports}\ }\textbf {\bibinfo {volume} {7}},\ \bibinfo {pages} {43365 (1--9)}
  (\bibinfo {year} {2017})},\ \Eprint {http://arxiv.org/abs/arXiv:1610.06164}
  {arXiv:1610.06164} \BibitemShut {NoStop}%
\bibitem [{\citenamefont {Auff\'eves}\ and\ \citenamefont
  {Grangier}(2018)}]{Auffeves-Grangier-2018}%
  \BibitemOpen
  \bibfield  {author} {\bibinfo {author} {\bibfnamefont {Alexia}\ \bibnamefont
  {Auff\'eves}}\ and\ \bibinfo {author} {\bibfnamefont {Philippe}\ \bibnamefont
  {Grangier}},\ }\bibfield  {title} {\enquote {\bibinfo {title}
  {Extracontextuality and extravalence in quantum mechanics},}\ }\href
  {\doibase 10.1098/rsta.2017.0311} {\bibfield  {journal} {\bibinfo  {journal}
  {Philosophical Transactions of the Royal Society {A}: Mathematical, Physical
  and Engineering Sciences}\ }\textbf {\bibinfo {volume} {376}},\ \bibinfo
  {pages} {20170311} (\bibinfo {year} {2018})}\BibitemShut {NoStop}%
\bibitem [{\citenamefont {Marcus}\ and\ \citenamefont
  {Ree}(1959)}]{Marcus-Ree-1959}%
  \BibitemOpen
  \bibfield  {author} {\bibinfo {author} {\bibfnamefont {M.}~\bibnamefont
  {Marcus}}\ and\ \bibinfo {author} {\bibfnamefont {R.}~\bibnamefont {Ree}},\
  }\bibfield  {title} {\enquote {\bibinfo {title} {{Diagonals of doubly
  stochastic matrices}},}\ }\href {\doibase 10.1093/qmath/10.1.296} {\bibfield
  {journal} {\bibinfo  {journal} {The Quarterly Journal of Mathematics}\
  }\textbf {\bibinfo {volume} {10}},\ \bibinfo {pages} {296--302} (\bibinfo
  {year} {1959})}\BibitemShut {NoStop}%
\bibitem [{\citenamefont {Mermin}(2007)}]{mermin-07}%
  \BibitemOpen
  \bibfield  {author} {\bibinfo {author} {\bibfnamefont {David~N.}\
  \bibnamefont {Mermin}},\ }\href {\doibase 10.1017/CBO9780511813870} {\emph
  {\bibinfo {title} {Quantum Computer Science}}}\ (\bibinfo  {publisher}
  {Cambridge University Press},\ \bibinfo {address} {Cambridge},\ \bibinfo
  {year} {2007})\BibitemShut {NoStop}%
\bibitem [{\citenamefont {Gr{\"o}tschel}\ \emph {et~al.}(1986)\citenamefont
  {Gr{\"o}tschel}, \citenamefont {Lov{\'a}sz},\ and\ \citenamefont
  {Schrijver}}]{GroetschelLovaszSchrijver1986}%
  \BibitemOpen
  \bibfield  {author} {\bibinfo {author} {\bibfnamefont {Martin}\ \bibnamefont
  {Gr{\"o}tschel}}, \bibinfo {author} {\bibfnamefont {L{\'a}szlo}\ \bibnamefont
  {Lov{\'a}sz}}, \ and\ \bibinfo {author} {\bibfnamefont {Alexander}\
  \bibnamefont {Schrijver}},\ }\bibfield  {title} {\enquote {\bibinfo {title}
  {Relaxations of vertex packing},}\ }\href {\doibase
  10.1016/0095-8956(86)90087-0} {\bibfield  {journal} {\bibinfo  {journal}
  {Journal of Combinatorial Theory, Series B}\ }\textbf {\bibinfo {volume}
  {40}},\ \bibinfo {pages} {330--343} (\bibinfo {year} {1986})}\BibitemShut
  {NoStop}%
\bibitem [{\citenamefont {Pitowsky}(2008)}]{Pitowsky-08-ge}%
  \BibitemOpen
  \bibfield  {author} {\bibinfo {author} {\bibfnamefont {Itamar}\ \bibnamefont
  {Pitowsky}},\ }\bibfield  {title} {\enquote {\bibinfo {title} {Geometry of
  quantum correlations},}\ }\href {\doibase 10.1103/PhysRevA.77.062109}
  {\bibfield  {journal} {\bibinfo  {journal} {Physical Review A}\ }\textbf
  {\bibinfo {volume} {77}},\ \bibinfo {pages} {062109} (\bibinfo {year}
  {2008})},\ \Eprint {http://arxiv.org/abs/arXiv:0802.3632} {arXiv:0802.3632}
  \BibitemShut {NoStop}%
\bibitem [{\citenamefont {Bell}(1990)}]{bell-a}%
  \BibitemOpen
  \bibfield  {author} {\bibinfo {author} {\bibfnamefont {John~Stuard}\
  \bibnamefont {Bell}},\ }\bibfield  {title} {\enquote {\bibinfo {title}
  {Against `measurement'},}\ }\href {\doibase 10.1088/2058-7058/3/8/26}
  {\bibfield  {journal} {\bibinfo  {journal} {Physics World}\ }\textbf
  {\bibinfo {volume} {3}},\ \bibinfo {pages} {33--41} (\bibinfo {year}
  {1990})}\BibitemShut {NoStop}%
\bibitem [{\citenamefont {Lakatos}(1978)}]{lakatosch}%
  \BibitemOpen
  \bibfield  {author} {\bibinfo {author} {\bibfnamefont {Imre}\ \bibnamefont
  {Lakatos}},\ }\href@noop {} {\emph {\bibinfo {title} {Philosophical Papers.
  1. {T}he Methodology of Scientific Research Programmes}}}\ (\bibinfo
  {publisher} {Cambridge University Press},\ \bibinfo {address} {Cambridge},\
  \bibinfo {year} {1978})\BibitemShut {NoStop}%
\bibitem [{\citenamefont {Cabello}\ \emph {et~al.}(2014)\citenamefont
  {Cabello}, \citenamefont {Severini},\ and\ \citenamefont
  {Winter}}]{Cabello-2014-gtatqc}%
  \BibitemOpen
  \bibfield  {author} {\bibinfo {author} {\bibfnamefont {Ad\'an}\ \bibnamefont
  {Cabello}}, \bibinfo {author} {\bibfnamefont {Simone}\ \bibnamefont
  {Severini}}, \ and\ \bibinfo {author} {\bibfnamefont {Andreas}\ \bibnamefont
  {Winter}},\ }\bibfield  {title} {\enquote {\bibinfo {title} {Graph-theoretic
  approach to quantum correlations},}\ }\href {\doibase
  10.1103/PhysRevLett.112.040401} {\bibfield  {journal} {\bibinfo  {journal}
  {Physical Review Letters}\ }\textbf {\bibinfo {volume} {112}},\ \bibinfo
  {pages} {040401} (\bibinfo {year} {2014})},\ \Eprint
  {http://arxiv.org/abs/arXiv:1401.7081} {arXiv:1401.7081} \BibitemShut
  {NoStop}%
\bibitem [{\citenamefont {Specker}(1990)}]{specker-ges}%
  \BibitemOpen
  \bibfield  {author} {\bibinfo {author} {\bibfnamefont {Ernst}\ \bibnamefont
  {Specker}},\ }\href {\doibase 10.1007/978-3-0348-9259-9} {\emph {\bibinfo
  {title} {Selecta}}}\ (\bibinfo  {publisher} {Birkh{\"{a}}user Verlag},\
  \bibinfo {address} {Basel},\ \bibinfo {year} {1990})\BibitemShut {NoStop}%
\bibitem [{\citenamefont {Zierler}\ and\ \citenamefont
  {Schlessinger}(1975)}]{Zierler1975}%
  \BibitemOpen
  \bibfield  {author} {\bibinfo {author} {\bibfnamefont {Neal}\ \bibnamefont
  {Zierler}}\ and\ \bibinfo {author} {\bibfnamefont {Michael}\ \bibnamefont
  {Schlessinger}},\ }\bibfield  {title} {\enquote {\bibinfo {title} {Boolean
  embeddings of orthomodular sets and quantum logic},}\ }in\ \href {\doibase
  10.1007/978-94-010-1795-4_14} {\emph {\bibinfo {booktitle} {The
  Logico-Algebraic Approach to Quantum Mechanics: Volume {I}: Historical
  Evolution}}},\ \bibinfo {editor} {edited by\ \bibinfo {editor} {\bibfnamefont
  {C.~A.}\ \bibnamefont {Hooker}}}\ (\bibinfo  {publisher} {Springer
  Netherlands},\ \bibinfo {address} {Dordrecht},\ \bibinfo {year} {1975})\ pp.\
  \bibinfo {pages} {247--262}\BibitemShut {NoStop}%
\end{thebibliography}

%

\end{document}